\documentclass{emulateapj}
\usepackage{amssymb,amsmath}
\usepackage{verbatim}
\usepackage{color,hyperref}
\usepackage{needspace}
\definecolor{linkcolor}{rgb}{0,0,0.25}
\hypersetup{
  colorlinks=true,        
  linkcolor=linkcolor,    
  citecolor=linkcolor,    
  filecolor=linkcolor,    
  urlcolor=linkcolor      
}
\newcounter{address}
\newcommand{\ie}{i.e.}
\newcommand{\etal}{et al.}

\newcommand{\eg}{e.g.}
\newcommand{\eqnname}{equation}
\newcommand{\Eqnname}{Equation}
\newcommand{\equationname}{\eqnname}

\renewcommand{\figurename}{Figure}

\newcommand{\sectionname}{$\mathsection$}

\newcommand{\feh}{\ensuremath{[\mathrm{Fe/H}]}}

\newcommand{\Ro}{\ensuremath{R_0}}

\newcommand{\Gyr}{\ensuremath{\,\mathrm{Gyr}}}
\newcommand{\kpc}{\ensuremath{\,\mathrm{kpc}}}
\newcommand{\pc}{\ensuremath{\,\mathrm{pc}}}
\newcommand{\kms}{\ensuremath{\,\mathrm{km\ s}^{-1}}}
\newcommand{\msun}{\ensuremath{\,\mathrm{M}_{\odot}}}

\newcommand{\inv}{\ensuremath{^{-1}}}

\newcommand{\magunit}{\,\mbox{mag}}
\newcommand{\ks}{\ensuremath{K_s}}

\newcommand{\logg}{\ensuremath{\log g}}
\newcommand{\teff}{\ensuremath{T_{\mathrm{eff}}}}

\newcommand{\vsun}{\ensuremath{V_{\odot-c}}}

\submitted{}

\begin{document}

\title{The power spectrum of the Milky Way: Velocity fluctuations in
  the Galactic disk}

\author{Jo~Bovy\altaffilmark{1,2,3},
  Jonathan~C.~Bird\altaffilmark{4,5},
  Ana~E.~Garc\'{\i}a~P\'{e}rez\altaffilmark{6,7}, 
  Steven~R.~Majewski\altaffilmark{6},
  David~L.~Nidever\altaffilmark{8},
  and Gail~Zasowski\altaffilmark{9}}
\altaffiltext{\theaddress}{\label{1}\stepcounter{address} 
  Institute for Advanced Study, Einstein Drive, Princeton, NJ 08540, USA; bovy@ias.edu~}
\altaffiltext{\theaddress}{\label{2}\stepcounter{address} 
  Hubble Fellow}
\altaffiltext{\theaddress}{\label{3}\stepcounter{address} 
  John Bahcall Fellow}
\altaffiltext{\theaddress}{\label{4}\stepcounter{address} 
  Department of Physics and Astronomy, Vanderbilt University, 6301 Stevenson Center, Nashville, TN, 37235, USA}
\altaffiltext{\theaddress}{\label{5}\stepcounter{address} 
  VIDA Postdoctoral Fellow}
\altaffiltext{\theaddress}{\label{6}\stepcounter{address} 
  Department of Astronomy, University of Virginia, Charlottesville, VA, 22904, USA}
\altaffiltext{\theaddress}{\label{7}\stepcounter{address} 
  Instituto de Astrof\'{\i}sica de Canarias, E38205 La Laguna, Tenerife, Spain}
\altaffiltext{\theaddress}{\label{9}\stepcounter{address} 
  Department of Astronomy, University of Michigan, Ann Arbor, MI, 48104, USA}
\altaffiltext{\theaddress}{\label{10}\stepcounter{address} 
  Department of Physics and Astronomy, Johns Hopkins University, Baltimore, MD 21218, USA}

\begin{abstract}  
  We investigate the kinematics of stars in the mid-plane of the Milky
  Way on scales between $25\pc$ and $10\kpc$ with data from the Apache
  Point Observatory Galactic Evolution Experiment (APOGEE), the Radial
  Velocity Experiment (RAVE), and the Geneva-Copenhagen Survey
  (GCS). Using red-clump stars in APOGEE, we determine the large-scale
  line-of-sight velocity field out to $5\kpc$ from the Sun in
  $(0.75\kpc)^2$ bins. The solar motion \vsun\ with respect to the
  circular velocity $V_c$ is the largest contribution to the power on
  large scales after subtracting an axisymmetric rotation field; we
  determine the solar motion by minimizing the large-scale power to be
  $\vsun = 24\pm 1\, (\mathrm{ran.})\pm 2\, (\mathrm{syst.}\ [V_c])\pm
  5\,(\mathrm{syst.\ [large\!-\!scale]})\kms$, where the systematic
  uncertainty is due to (a) a conservative $20\kms$ uncertainty in
  $V_c$ and (b) the estimated power on unobserved larger
  scales. Combining the APOGEE peculiar-velocity field with red-clump
  stars in RAVE out to $2\kpc$ from the Sun and with local GCS stars,
  we determine the power spectrum of residual velocity fluctuations in
  the Milky Way's disk on scales between $0.2\kpc\inv \leq k \leq
  40\kpc\inv$. Most of the power is contained in a broad peak between
  $0.2\kpc\inv < k < 0.9\kpc\inv$. We investigate the expected power
  spectrum for various non-axisymmetric perturbations and demonstrate
  that the central bar with commonly used parameters but of relatively
  high mass can explain the bulk of velocity fluctuations in the plane
  of the Galactic disk near the Sun. Streaming motions $\approx10\kms$
  on $\gtrsim3\kpc$ scales in the Milky Way are in good agreement with
  observations of external galaxies and directly explain why local
  determinations of the solar motion are inconsistent with global
  measurements.
\end{abstract}

\keywords{
        Galaxy: disk
        ---
        Galaxy: fundamental parameters
        ---
        Galaxy: general
        ---        
        Galaxy: kinematics and dynamics
        ---
        Galaxy: structure
        ---
        stars: kinematics
}

\section{Introduction}

One of the major questions in galactic astrophysics is what drives the
evolution of large disk galaxies such as the Milky Way (MW). In
particular, finding the source of the dependence of the kinematics of
stars on age or elemental abundances has been surprisingly
difficult. \citet{Spitzer51a} presciently suggested that giant
molecular clouds can increase the random velocities of stars, while
\citet{Barbanis67a} proposed that spiral structure is responsible for
this effect. Other proposals include that kinematic heating occurs
through encounters with massive black holes \citep{Lacey85a} or
satellite galaxies \citep{Toth92a,Quinn93a,Velazquez99a}, or that the
velocity dispersion at birth of (at least some part of) the disk was
high \citep[\eg,][]{Brook04a,Bournaud09a,Bird13a,Stinson13}. These
various mechanisms have so far primarily been tested using
observations of the age-dependence of the velocity dispersion of stars
in the solar neighborhood \citep[\eg,][]{Wielen77a,Nordstroem04a} and
the ratio of the vertical to the radial velocity dispersion
$\sigma_z/\sigma_R$. While the observed density of molecular clouds is
too small to produce the observed heating rate and spiral structure is
inefficient at generating large vertical velocities
\citep{Carlberg87a,Binney88a}, the combination of the two makes
predictions that are largely consistent with observations of young and
intermediate-age stars. This is because transient spiral structure is
efficient at developing large radial velocity dispersions at the
observed rate \citep{Carlberg85a} and scattering of molecular clouds
is highly efficient at keeping $\sigma_z/\sigma_R$ constant at
$\approx0.5$ to $0.6$ \citep{Binney88a,Ida93a,Hanninen02a}. However,
strong positive evidence for this picture is currently lacking.

Besides spiral structure, the Milky Way is known to host a central bar
\citep{Blitz91a,Binney91a} with a mass of about $10^{10}\msun$
\citep{Zhao96a,Weiner99a} and a semimajor axis of about $3\kpc$
\citep{Binney97a,Bissantz02a}. The impact of the bar on the kinematics
of the stellar disk is typically expected to be small, except near the
bar's ends or close to the outer Lindblad resonance, unless the
pattern speed changes significantly. The latter is unlikely, as the
pattern speed would typically decrease due to dynamical friction
\citep{Weinberg85a} and the Milky Way's bar is almost as fast as it
can be given the bar's length (bar corotation should be beyond the
bar's semimajor axis; \citealt{Contopoulos80a}), possibly indicating
that it has not significantly slowed down. However, the bar may have
grown and slowed down simultaneously, thus remaining fast in the above
sense \citep{Athanassoula03a}. That the Milky Way's disk is maximal
\citep{Bovy13a} also predicts that the bar maintains a large pattern
speed for many dynamical times \citep{Debattista00a}. However,
depending on the relative location of the bar and spiral resonances,
the bar could have a large impact on the heating and migration rate of
stars near the Sun \citep{Minchev10a}, in particular leading to
stronger radial migration than would be produced by spiral structure
alone \citep{Sellwood02a}. Constraining the properties of the bar and
the spiral structure is therefore essential for improving our
understanding of the evolution of the stellar disk.

\emph{Hipparcos} \citep{ESA97a} and more recently the RAVE survey
\citep{Steinmetz06a} have added additional constraints to this picture
through observations of the velocity field near the
Sun. \citet{Dehnen98a} reconstructed the full local three-dimensional
velocity distribution function (DF) in the solar neighborhood from
astrometric \emph{Hipparcos} data, revealing a distribution that is
far from smooth and is characterized by half a dozen large
overdensities or moving groups. Subsequent investigations demonstrated
that these systems are likely due to dynamical effects rather than
incomplete mixing of newly-formed stars
\citep[\eg,][]{Bensby07a,Famaey08a,Bovy10b,Sellwood10a}. In
particular, \citet{DeSimone04a} established that the same type of
transient spiral structure that is consistent with the observed
heating rate can generate clumps in the local velocity
DF. \citet{Dehnen00a} further demonstrated that the action of the
central bar on the local stellar disk can create a moving group
similar to the Hercules moving group if the Sun is near the bar's
outer Lindblad resonance. More recently, \citet{Siebert11a} and
\citet{Williams13a} have performed measurements of the mean velocity
field of kinematically-warm stars using RAVE data; they detect a
gradient of $\approx -4\kms\kpc\inv$ in the Galactocentric radial
velocity and a complicated pattern of fluctuations not easily
explained through a single perturber (but see
\citealt{Faure14a}). Another detection of a non-axisymmetric velocity
was reported by \citet{BovyVc} (hereafter B12), who found that the
solar neighborhood is traveling $\sim\!14\kms$ ahead of the circular
velocity in their axisymmetric model.

A particular aspect of non-axisymmetric streaming motions is that they
may affect determinations of the Milky Way's circular velocity $V_c$
and of the Sun's motion relative to $V_c$. The sum of these---the
Sun's Galactocentric rotational velocity---is crucial for converting
observed kinematics from the heliocentric frame to the Galactocentric
frame, which is the natural frame for interpreting large-scale
kinematics. The precisely measured proper motion of Sgr A$^*$
\citep{Reid04a} provides a crucial constraint on the Sun's
Galactocentric velocity (assuming that Sgr A$^*$ is at rest with
respect to the dynamical center of the disk and halo), but the present
uncertainty in the value of the distance to the Galactic center means
that the uncertainty in the heliocentric-to-Galactocentric coordinate
transformation is a serious limiting factor in many analyses of the
Milky Way. Various determinations of the Sun's motion with respect to
$V_c$ based on solar-neighborhood observations
\citep[\eg,][]{Dehnen98a,Schoenrich10a} and larger-scale observations
(\eg, B12) differ at the many-sigma level based on their formal
uncertainties. This may be due to the unaccounted influence of
non-axisymmetry on the kinematics of solar-neighborhood stars. 

In this paper we address these issues by exploring the large-scale
kinematics of red-clump (RC) stars in the MW disk from the APOGEE
survey (\citealt{Eisenstein11a}; S.~R.~Majewski \etal\ 2014, in
preparation). The first part of the paper focuses on investigating the
large solar motion---defined here as the relative velocity between the
Sun and the circular velocity at the solar circle in an axisymmetric
approximation of the Milky Way and denoted by \vsun\footnote{A perhaps
  more traditional definition of the solar motion is the motion of the
  Sun relative to the velocity of a hypothetical stellar population
  with zero velocity dispersion, which we denote as $V_\odot$. If the
  Milky Way is axisymmetric then these two definitions agree and
  $V_\odot \equiv \vsun$, but more generally they can be
  different.}---found by B12 but by instead using the RC sample, which
is superior to the sample used by B12 because it has much more
accurate distances. We find strong evidence in the RC kinematics that
the solar motion in the direction of rotation is $\vsun\sim\!24\kms$
and therefore that the solar neighborhood is indeed moving
$\sim\!12\kms$ faster than the circular velocity. In the second part
of the paper, we investigate this result by studying the mean velocity
field's deviations from axisymmetric rotation on scales ranging from
$25\pc$ to $10\kpc$. We calculate the power spectrum of fluctuations
and find that it exhibits a clear peak at scales corresponding to the
response to the central bar.

The outline of this paper is as follows. In
\sectionname~\ref{sec:data} we briefly discuss the various sources of
data we employ to study the velocity field on different scales, and
\sectionname~\ref{sec:psdmethod} contains our conventions for
calculating the two- and one-dimensional power spectrum of velocity
fluctuations. We investigate the dependence of the APOGEE-RC
kinematics on the assumed solar motion and determine a new value for
the solar motion by minimizing the large-scale power in the RC
kinematics in \sectionname~\ref{sec:solarmotion}. The observed
velocity fluctuations on different scales and their power spectrum are
presented in \sectionname~\ref{sec:obspsd}. We investigate the
expected power spectrum for various types of non-axisymmetric
perturbations in \sectionname~\ref{sec:modelpsd} and identify the
central bar as the most plausible agent responsible for the observed
fluctuations.  We discuss our results in the context of previous
findings in \sectionname~\ref{sec:discussion} and present our
conclusions in \sectionname~\ref{sec:conclusion}. We fix the Sun's
distance to the Galactic center to $\Ro = 8\kpc$ throughout our
analysis. In this study we calculate Galactocentric coordinates
$(X_{\mathrm{GC}},Y_{\mathrm{GC}},Z)$ in a left-handed frame with the
Sun at $(X_{\mathrm{GC}},Y_{\mathrm{GC}}) = (8,0)\kpc$,
$X_{\mathrm{GC}}$ positive toward $l=180^\circ$, $Y_{\mathrm{GC}}$
increasing toward Galactic rotation, and $Z$ positive toward the North
Galactic Pole. In all of the figures, the disk rotates
counter-clockwise, that is, it is as seen from the South Galactic
Pole.

\section{Data}\label{sec:data}

\begin{figure*}[t!]
\includegraphics[width=\textwidth,clip=]{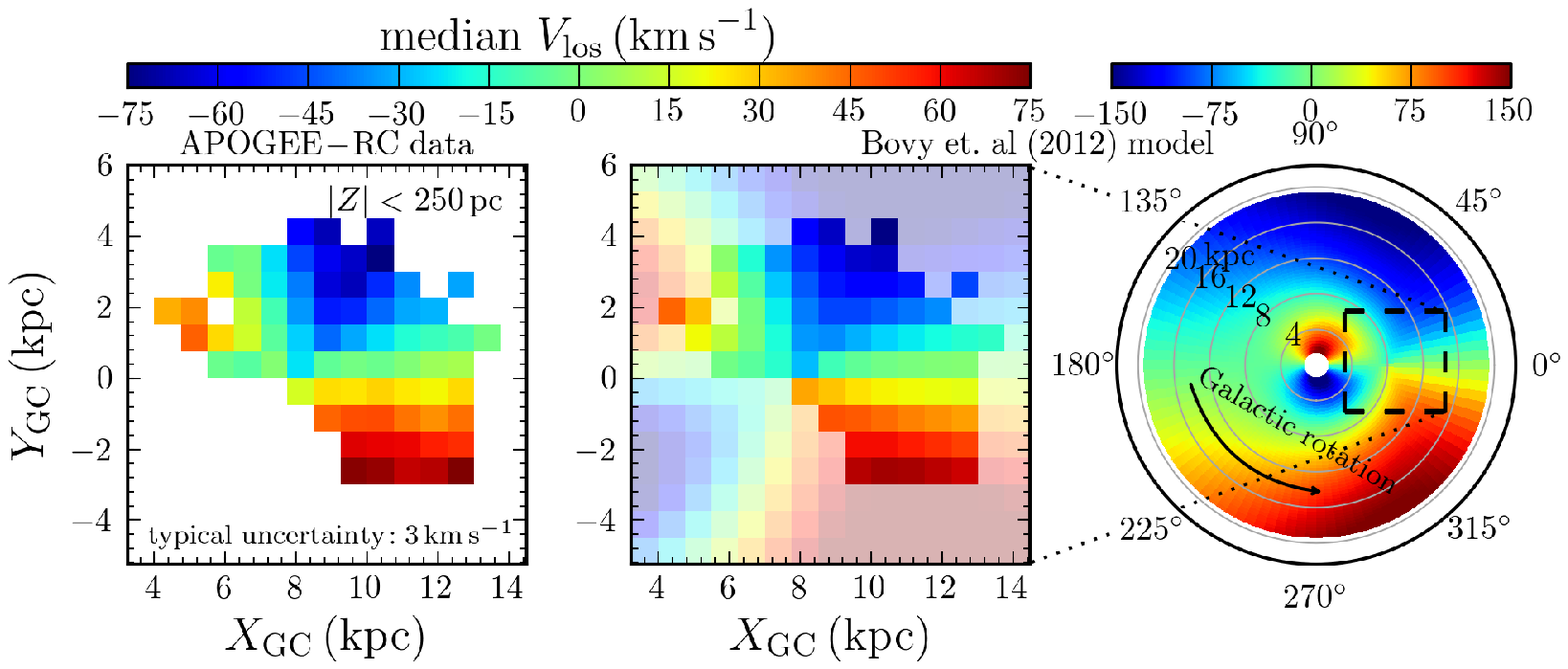}
\caption{Line-of-sight velocity field in the MW out to $5\kpc$ from
  the Sun from APOGEE red-clump stars close to the mid-plane
  (\emph{left panel}). This figure displays the median heliocentric
  line-of-sight velocity in $(0.75\kpc)^2$ pixels in rectangular,
  Galactocentric coordinates. The right panel displays the
  line-of-sight velocity field over the whole disk obtained in the B12
  model for the Milky-Way rotation curve and for the DF of RC
  stars. The middle panel is an expanded view of the model prediction
  in the APOGEE volume, with the areas of spatial overlap with APOGEE
  data highlighted. The velocity color range on the right is twice
  that in the left and middle panels. The Sun is located at
  $(X_{\mathrm{GC}},Y_{\mathrm{GC}}) = (8,0)\kpc$.}\label{fig:rcvlos}
\end{figure*}

We primarily use data from the APOGEE survey, but also from RAVE and
the Geneva-Copenhagen survey (GCS; \citealt{Nordstroem04a}) to
investigate velocity fluctuations on smaller scales than is possible
with APOGEE. We describe all three data sets briefly in this section.

We use data on the kinematics of RC stars from the DR12 APOGEE-RC
catalog, which is constructed using the method described in detail in
\citet{BovyRC}, but applied to the DR12 data. Briefly, RC stars are
selected from the superset of all APOGEE data in SDSS-III DR12 (C.~Ahn
\etal\ 2015, in preparation) using a combination of cuts in the
surface-gravity (\logg) / effective-temperature (\teff) plane and the
$(J-\ks)_0$ / metallicity (\feh) plane. These cuts are chosen to
select a pure sample of RC stars for which precise luminosity
distances can be determined. Reddening and extinction corrections are
determined for each star using the Rayleigh Jeans Color Excess method
\citep{Majewski11a}; for the RC, random and systematic uncertainties
in these corrections are $\lesssim0.05\magunit$ (see
\citealt{BovyRC}). The distance scale is calibrated against an
\emph{Hipparcos}-based determination of the RC absolute $\ks$
magnitude in the solar neighborhood \citep{Laney12a}. Corrections to a
single absolute magnitude as a function of $([J-\ks]_0,\feh)$ based on
stellar isochrone models \citep{Bressan12a} are determined for each
star in the catalog individually. The catalog has an estimated purity
of $\approx 95\,\%$ and the distances are precise to $5\,\%$ and
unbiased to $2\,\%$. The uncertainty in the line-of-sight velocities
is typically $0.1\kms$. Full details on the APOGEE survey can be found
in S.~R.~Majewski \etal\ (2015, in preparation), on the APOGEE
instrument in \citet{Wilson10a} and J.~Wilson \etal\ (2015, in
preparation), on the Sloan 2.5-meter telescope in \citet{Gunn06a}, on
the APOGEE target selection in \citet{Zasowski13a}, on the data
reduction pipeline in D.~L.~Nidever \etal\ (2015, in preparation), on
the stellar-parameters and chemical-abundances analysis in
A.~E.~Garc\'{\i}a~P\'{e}rez \etal\ (2015, in preparation), and on the
specific DR12 analysis and calibration in J.~Holtzman \etal\ (2015, in
preparation).

\begin{figure*}[t!]
\begin{center}
\includegraphics[width=0.45\textwidth,clip=]{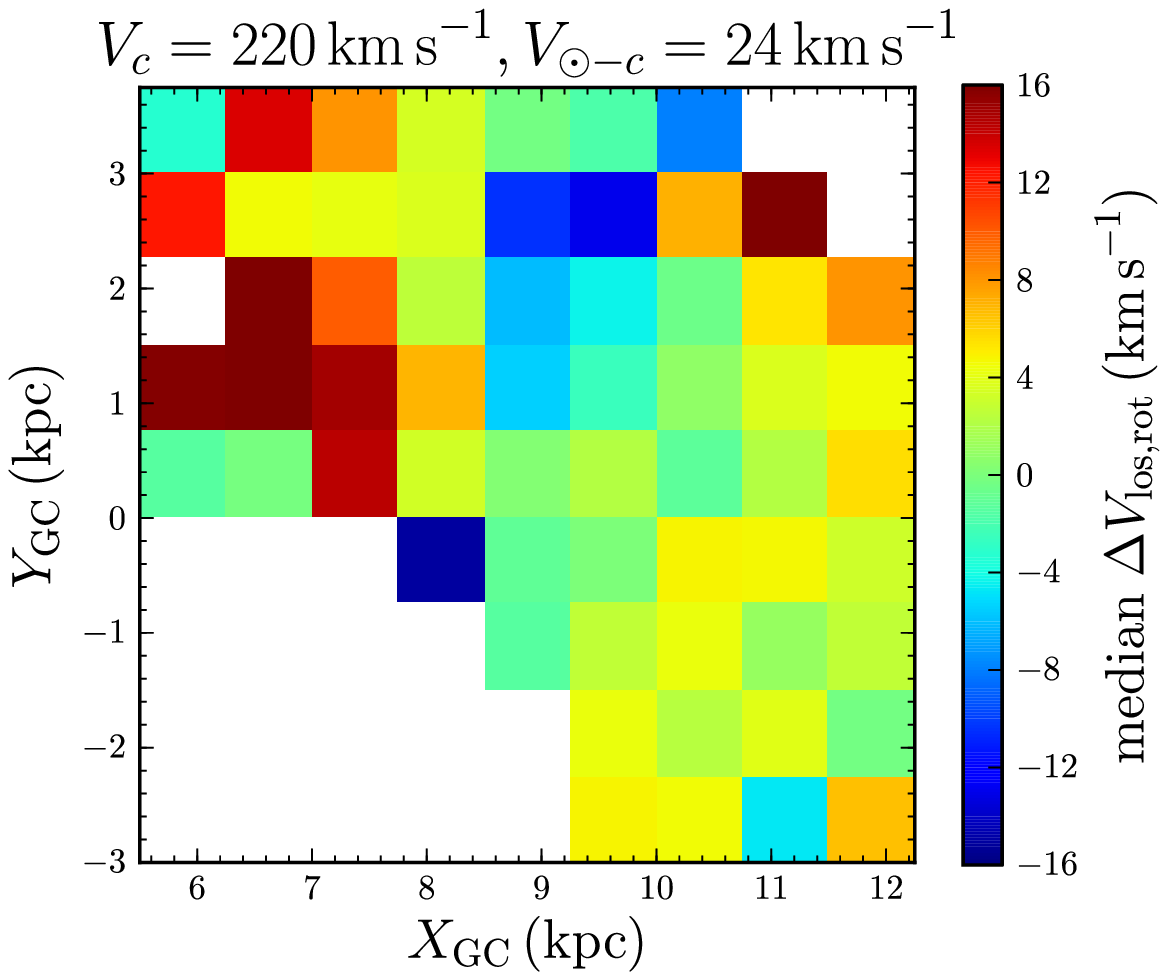}
\includegraphics[width=0.45\textwidth,clip=]{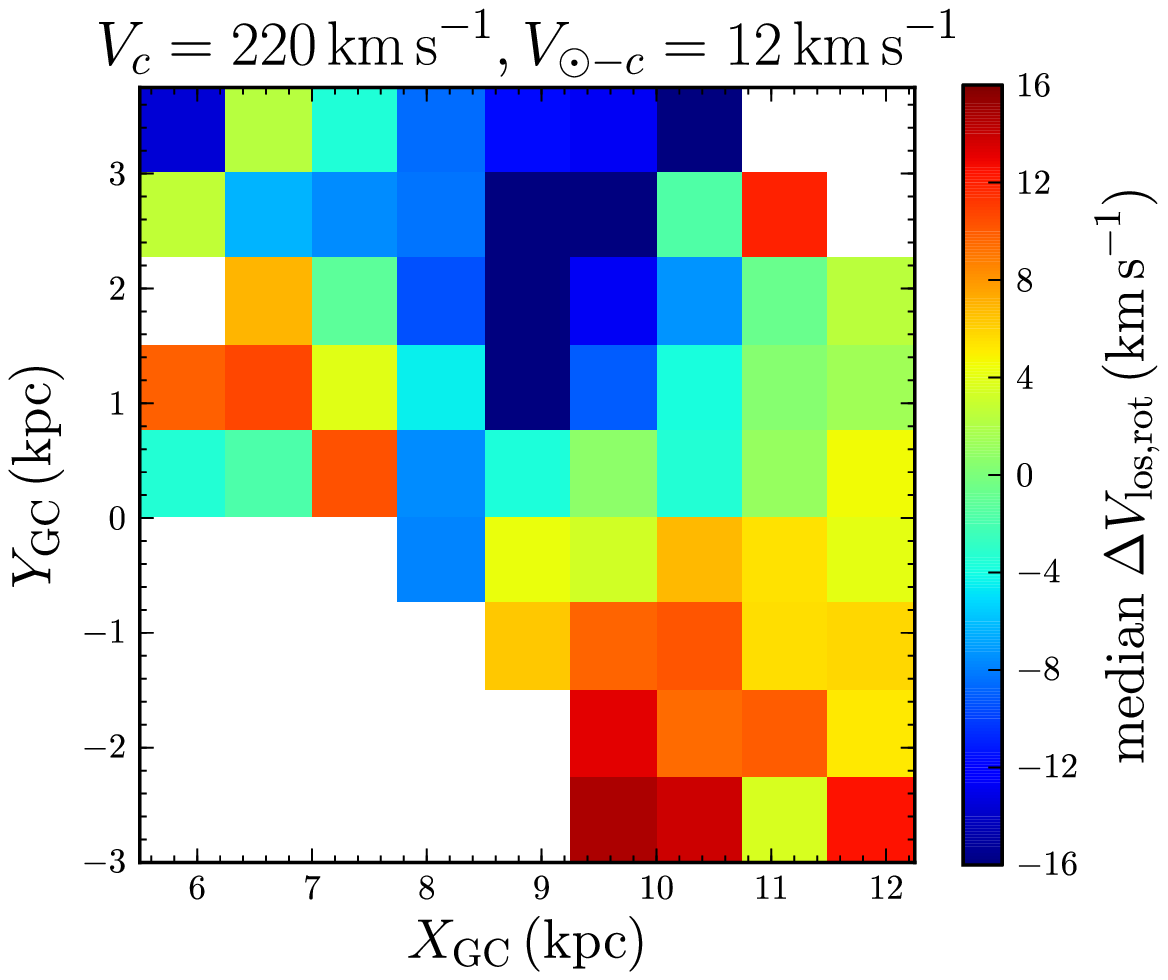}\\
\includegraphics[width=0.45\textwidth,clip=]{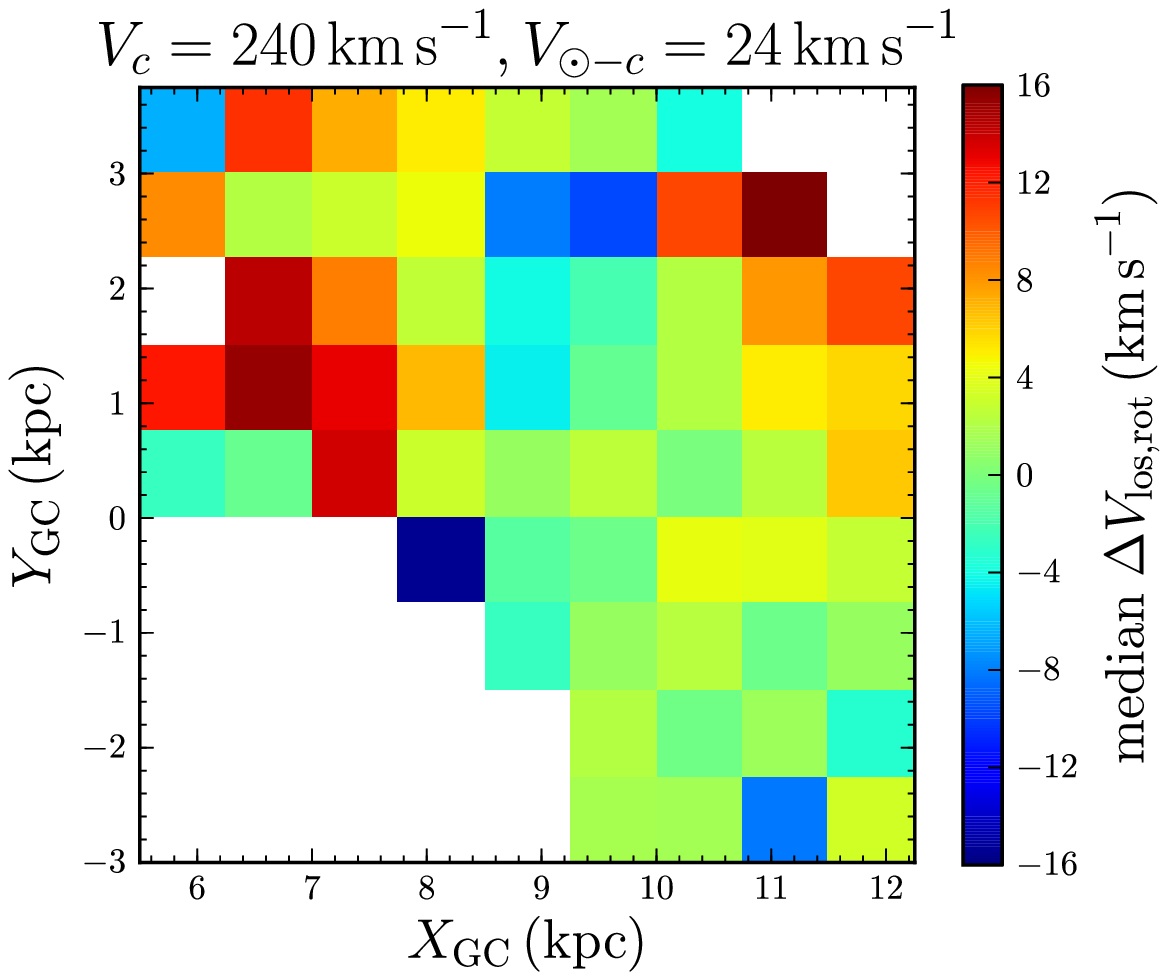}
\includegraphics[width=0.45\textwidth,clip=]{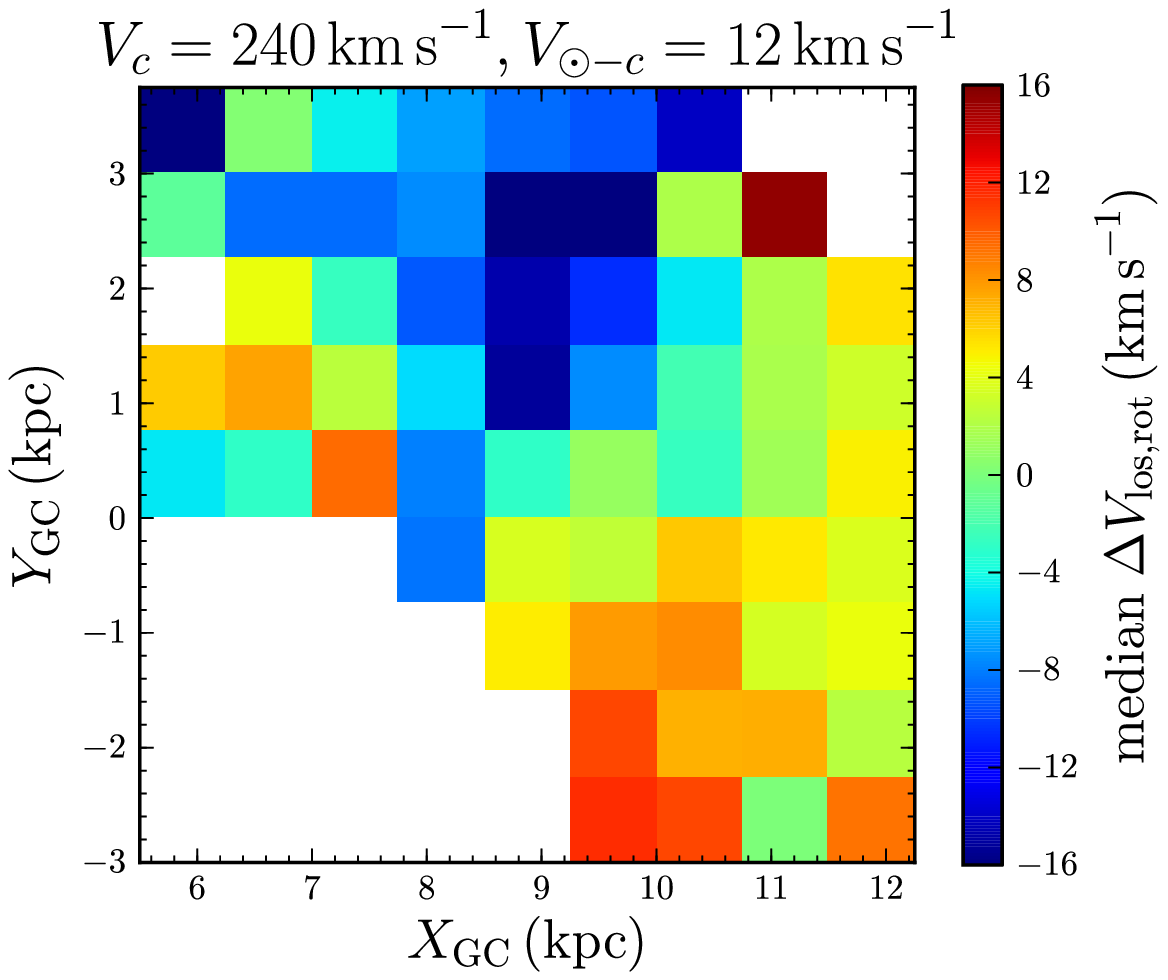}
\end{center}
\caption{Peculiar velocity field in the Milky Way out to $5\kpc$ from
  the Sun. These figures display the residuals between the median
  heliocentric line-of-sight velocities in
  \figurename~\ref{fig:rcvlos} and the model of B12
  (\equationname~[\ref{eq:deltav}]), for different values of the
  circular velocity $V_c$ and the solar motion $\vsun$. The peculiar
  velocity field depends sensitively on the assumed solar motion, but
  is largely independent of the assumed $V_c$.}\label{fig:rcdvlos}
\end{figure*}

From the APOGEE-RC catalog we select 8,155 stars within $250\pc$ from
the Galactic mid-plane. The median heliocentric line-of-sight
velocities $V_{\mathrm{los}}$ of this sample are displayed in the left
panel of \figurename~\ref{fig:rcvlos}, which presents the median of
the line-of-sight velocities in $(0.75\kpc)^2$ pixels. All pixels here
and in the following figures have a minimum of 20 stars per pixel. The
typical uncertainty on the median velocity is $3\kms$, with smaller
values near the Sun and larger values at the far edges of the
sample. The observed heliocentric line-of-sight velocity field
reflects the combination of the rotation of the Galactic disk and the
motion of the Sun with respect to this rotation (see, \eg, B12) and
spans a range of $\approx 150\kms$ in the observed volume. With a
typical uncertainty of $3\kms$, we therefore measure this combination
at extremely high signal-to-noise ratio. The right panel of
\figurename~\ref{fig:rcvlos} displays the line-of-sight velocity field
over the whole Galactic disk obtained for the model of B12, which was
fit to the kinematics of stars observed during the first year of
APOGEE. The middle panel is an expanded view of the region observed by
the APOGEE-RC sample. The residuals between the data and the B12 model
are displayed in the top-left panel of
\figurename~\ref{fig:rcdvlos}. The overall agreement between the data
and the B12 model is excellent.

We also employ data from the RAVE survey, selecting a sample of RC
stars in RAVE by applying almost the same cuts\footnote{We do
  \emph{not} apply the additional cut in \logg\ and
  \teff\ (\equationname~[9] in \citealt{BovyRC}) to RAVE that was
  necessary for APOGEE because of a relative \logg\ bias for RC and
  red-giant stars (see \sectionname~2.3 of \citealt{BovyRC}).} applied
to create the APOGEE-RC sample to the DR4 RAVE data set
\citep{Kordopatis13a}, using the stellar parameters determined by
RAVE's Stellar Parameter Pipeline and dereddening the $J-\ks$ color
using $E_{J-\ks} = 0.17\,A_V$. We do not perform any other cuts. This
approach creates a sample of 30,783 RC stars in RAVE, which have
spectra with a typical signal-to-noise ratio of 50 per resolution
element, line-of-sight velocities with uncertainties of typically
$1\kms$, and which are typically within $\approx2\kpc$ from the
Sun. For these stars we use the distances from \citet{Binney14a}.

Data on the velocity field on small scales is obtained from the GCS
catalog \citep{Nordstroem04a}. We select stars from the GCS catalog
that are not flagged as binaries and that have a measured metallicity
larger than $-1.2$; 8,123 stars in GCS satisfy these
constraints. While for RAVE and APOGEE we use the line-of-sight
velocity to investigate the kinematics, for GCS we use the $V$
component of the velocity vector, that is, the $y$ component of the
velocity field in the rectangular Galactic coordinate frame. We employ
the $V$ component rather than the $x$ component of the velocity field,
because it has a smaller dispersion, thus making it easier to
determine the median velocity.

\section{Two- and one-dimensional power spectrum}\label{sec:psdmethod}

\begin{figure}[t!]
\includegraphics[width=0.48\textwidth,clip=]{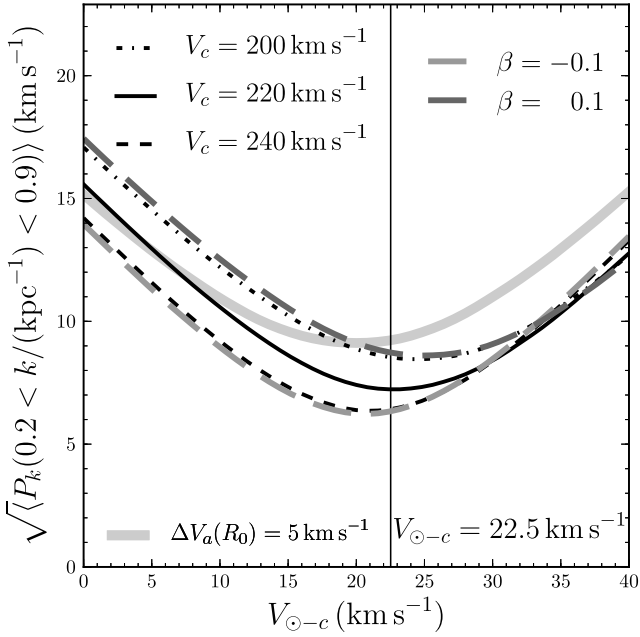}
\caption{Determination of the solar motion by minimizing the
  large-scale power in the peculiar-velocity field out to
  $5\kpc$. This figure presents the power on scales $0.2\kpc^{-1} < k
  < 0.9\kpc^{-1}$ as a function of the solar motion $\vsun$ for
  different models for the MW rotation curve. The power is minimized
  for the fiducial model of a flat rotation curve with $V_c = 220\kms$
  for a solar motion of $\vsun = 22.5\kms$. Assuming different $V_c$,
  a falling or rising rotation curve (parameterized as $V_c(R) \propto
  R^\beta$ here), or a different asymmetric-drift model (thick gray
  line, see text) only shifts this minimum by a few \kms. Simulations
  indicate that this determination of $\vsun$ is biased low by
  $\approx1\kms$ and that it has a random uncertainty of $1\kms$;
  therefore, our value for $\vsun$ is $\vsun =
  24\pm1\kms$.}\label{fig:vsolar}
\end{figure}

In \sectionname~\ref{sec:psd} we calculate the power spectrum of the
mean velocity field in the Milky Way after subtracting a smooth
axisymmetric model. The power spectrum is a simple, model-independent
transformation of the data that, as we argue below, highlights the
underlying physical mechanism responsible for non-axisymmetric motions
in a powerful manner. We determine the two-dimensional power spectrum
and also average the power spectrum azimuthally to determine the
one-dimensional power spectrum. As Fourier-transform and
power-spectrum conventions are not fully standardized, we briefly
discuss our procedure for performing these calculations.

For a two-dimensional field $a_{ij}$ on a rectangular $N\times M$ grid
$(x_i,y_j)$ with spacings ($\Delta_x$,$\Delta_y$), we start by padding
each $x$ and each $y$ dimension with $N+1$ and $M+1$ zeros,
respectively, to avoid periodic pollution in Fourier space
\citep{Press07a}. We then calculate the two-dimensional Fourier
transformation $A_{kl}$ as
\begin{equation}
  A_{kl} = \sum_{i=0}^{2N} \sum_{j=0}^{2M} a_{ij}\,\exp\left(-\pi I\left[\frac{ik}{N}+\frac{jl}{M}\right]\right)\,.
\end{equation}
Here, $I = \sqrt{-1}$. We then form the periodogram estimate of the
two-dimensional power spectrum $P(k_x,k_y)$ \citep[\eg,][]{Press07a}
at frequencies ($k_x,k_y$) = $(\alpha/N\,f_x,\beta/M\,f_y)$, where
$\alpha = 0,1,\ldots, N$ and $\beta = 0,1,\ldots,M$. We define $f_x =
1/\Delta_x$ and $f_y = 1/\Delta_y$. This definition is a factor of two
larger than that of the Nyquist frequency, but it allows the
approximate scale of a fluctuation to be determined as $k^{-1}$.

We construct the azimuthally-averaged, one-dimensional power spectrum
$P(k)$ by averaging $P(k_x,k_y)$ in bins of $k = \sqrt{k_x^2+k_y^2}$
and multiplying by $(4\pi)^2$
\begin{equation}
  P(k) = \frac{(4\pi)^2}{N_k}\,\sum_{(k_x,k_y)\ \mathrm{in\ bin}\ k} P(k_x,k_y)\,,
\end{equation}
where $N_k$ is the number of points in the $(k_x,k_y)$ grid that fall
within the $k$ bin.

\needspace{8ex}
\section{A novel, global determination of the solar motion}\label{sec:solarmotion}

\begin{figure*}[t!]
\includegraphics[width=\textwidth,clip=]{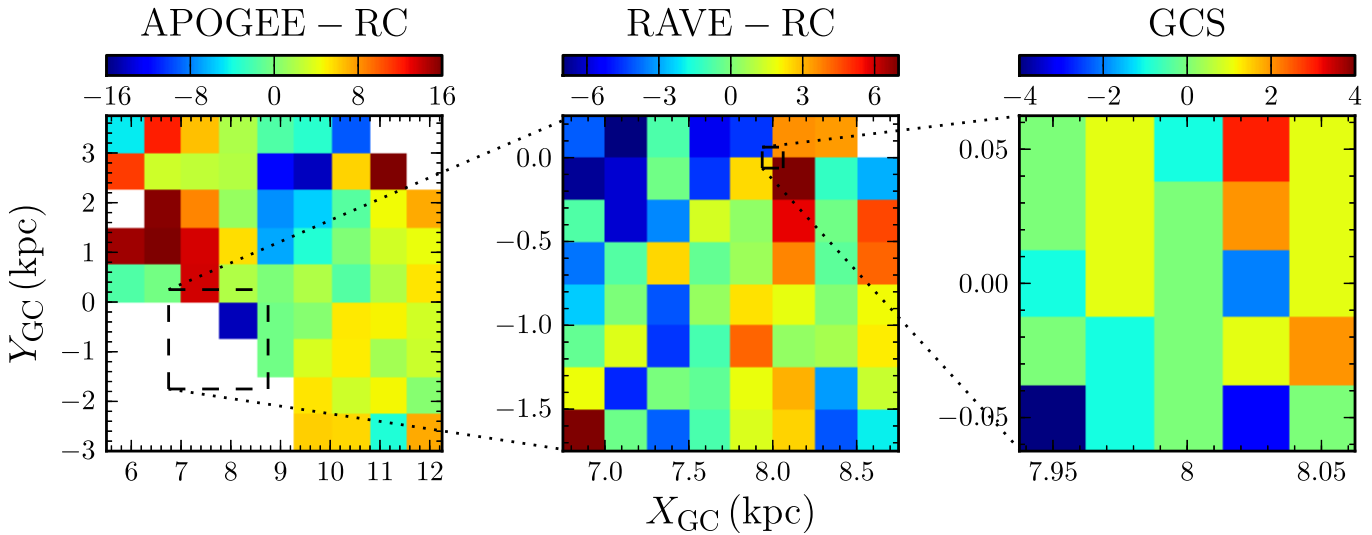}
\caption{Peculiar velocity field on different scales in the MW. This
  figure displays the MW's peculiar-velocity field in \kms\ on scales
  ranging from $25\pc$ to $10\kpc$ by combining three different
  surveys. The rightmost panel shows the fluctuations in the $V$
  component of the velocity of GCS stars, in $(25\pc)^2$ pixels in a
  box of size $(125\pc)^2$ around the Sun. The middle panel displays
  the peculiar velocity field of RAVE red-clump stars, in $(250\pc)^2$
  pixels covering a square of size $(2\kpc)^2$ around the Sun. The
  leftmost panel presents the peculiar velocity field in
  $(0.75\kpc)^2$ pixels out to $5\kpc$ (the same as the top, left
  panel of \figurename~\ref{fig:rcdvlos}, but using $\vsun =
  22.5\kms$ to minimize the large-scale power). The color scale ranges
  over a smaller range in velocities in the middle and right
  panels. There is a clear increase in the amplitude of the velocity
  fluctuations from the smallest to the largest
  scales.}\label{fig:2dvel}
\end{figure*}

The two-dimensional velocity field of heliocentric line-of-sight
velocities in \figurename~\ref{fig:rcvlos} can be modeled as arising
from the combination of three elements: (a) the overall differential
rotation of the Galactic disk (\ie, the circular-velocity curve), (b)
the offset between the mean velocity of the population of RC stars and
circular rotation (\ie, the asymmetric drift), and (c) the Sun's
motion with respect to the Galactocentric restframe. We do not attempt
to fully model all of these ingredients here; instead, we adopt the
best-fit model from B12, which consists of a model for the MW
potential (a simple power-law rotation curve) and for the stellar DF,
assumed to be axisymmetric and well-mixed. This model was fit to the
kinematics of a sample of stars at fourteen $b=0^\circ$ fields in the
first year of APOGEE data, by approximating the two-dimensional DF of
these stars as a bivariate Gaussian with a mean rotational-velocity
offset from circular rotation given by a simple asymmetric-drift model
and constant radial anisotropy.

The model in B12 was fit to all types of stars in APOGEE---not only RC
stars, but also first-ascent red giants and dwarfs. Only 782 of the
8,155 stars that we select here from the APOGEE-RC sample were part of
the sample of 3,365 stars used in the B12 analysis. Therefore, the
samples are largely independent. The B12 sample extends to larger
distances, because it includes luminous red giants up to $(J-\ks)_0 =
1.1$. For this reason the B12 sample was more suited for determining
the parameters of the circular velocity curve and in particular for
the circular velocity at the Sun itself. We thus keep most of the
parameters of the B12 model fixed for the RC sample used in this paper
and focus our attention on the solar motion in the rotational
direction. We set the radial-velocity dispersion of the Gaussian DF
model to $31.4\kms$, the radial scale length of the population to
$3\kpc$, and we fix the velocity-dispersion profile to be flat. The
Sun's radial motion $V_{R,\odot}$ is fixed to $-10.5\kms$ and the
Sun's vertical velocity $V_{Z,\odot}$ is set to $7.25\kms$
\citep{Schoenrich10a}.

With these parameter selections we can compute the expected mean
velocity at the location of each RC star in our sample. We calculate
the peculiar velocities $\Delta V_{\mathrm{los,rot}}$ with respect to
the mean model prediction at a point with Galactocentric cylindrical
coordinates $(R,\phi)$
\begin{equation}\label{eq:deltav}
\begin{split}
  \Delta  V_{\mathrm{los,rot}}(R,\phi) & = V_{\mathrm{los}} / \cos b - V_{R,\odot}\,\cos l \\ & \ + \left[V_c(R_0) + \vsun\right]\,\sin l + V_{Z,\odot}\,\sin b / \cos b\\
& \ - \left[V_c(R)-V_a(R)\right]\,\sin(l+\phi)\,,
\end{split}
\end{equation}
where $V_a(R)$ is the asymmetric drift. By taking the median of these
values we can determine the peculiar velocity field in the MW disk,
which should be zero if the disk is axisymmetric.

The peculiar velocity field calculated assuming a flat rotation curve
with $V_c = 220\kms$ and a solar motion $\vsun = 24\kms$ is displayed
in the top left panel of \figurename~\ref{fig:rcdvlos}. The deviations
from zero are large ($\sigma\approx7\kms$), but on average the
deviation is close to zero. The top right panel demonstrates the
velocity field using what is currently considered as the standard
solar motion, $\vsun = 12\kms$ (determined from solar neighborhood
kinematics and therefore assuming that $\vsun\equiv V_\odot$;
\citealt{Schoenrich10a}). In this case, a large-scale systematic trend
is visible in the peculiar velocity field. The bottom panels of
\figurename~\ref{fig:rcdvlos} illustrate that this conclusion is
largely independent of the assumed value for $V_c$: for $V_c =
240\kms$, similar patterns exist in the peculiar velocity field. This
is the case because the mean line-of-sight velocity field
$\bar{V}_{\mathrm{los}}$ assuming axisymmetry is given by
\begin{equation}
\bar{V}_{\mathrm{los}} = [V_c(R)-V_a(R)]\,\sin(\phi+l) -
    [V_c(R_0)+\vsun]\,\sin l\,,
\end{equation}
where for simplicity we have assumed that the Sun's radial motion is
zero. For a close-to-flat rotation curve ($V_c(R) \approx V_c(R_0)$)
and a narrow range in $\phi$, this expression reduces to
\begin{equation}
\bar{V}_{\mathrm{los}} \approx [V_c-V_a(R)]\,\sin\phi\cos l - [V_a(R)+\vsun]\,\sin l\,,
\end{equation}
The dependence on $V_c$ is $\propto \sin\phi\cos l$ and because of the
narrow $\phi$ range is therefore quite weak (typical $|\phi| \approx
12^\circ$, so typical $|\sin\phi| \approx 0.2$) while the dependence
on $\vsun$ is $\propto \sin l$ and thus strong (we vary both $V_c$ and
$\vsun$ over $40\kms$). As we discuss in
\sectionname~\ref{sec:discussion-vsolar} below, the difference in the
inferred $\vsun$ value here and that based on more local observations
can be explained by the difference in the survey volumes used in the
measurements.

\begin{figure*}[t!]
\includegraphics[width=\textwidth,clip=]{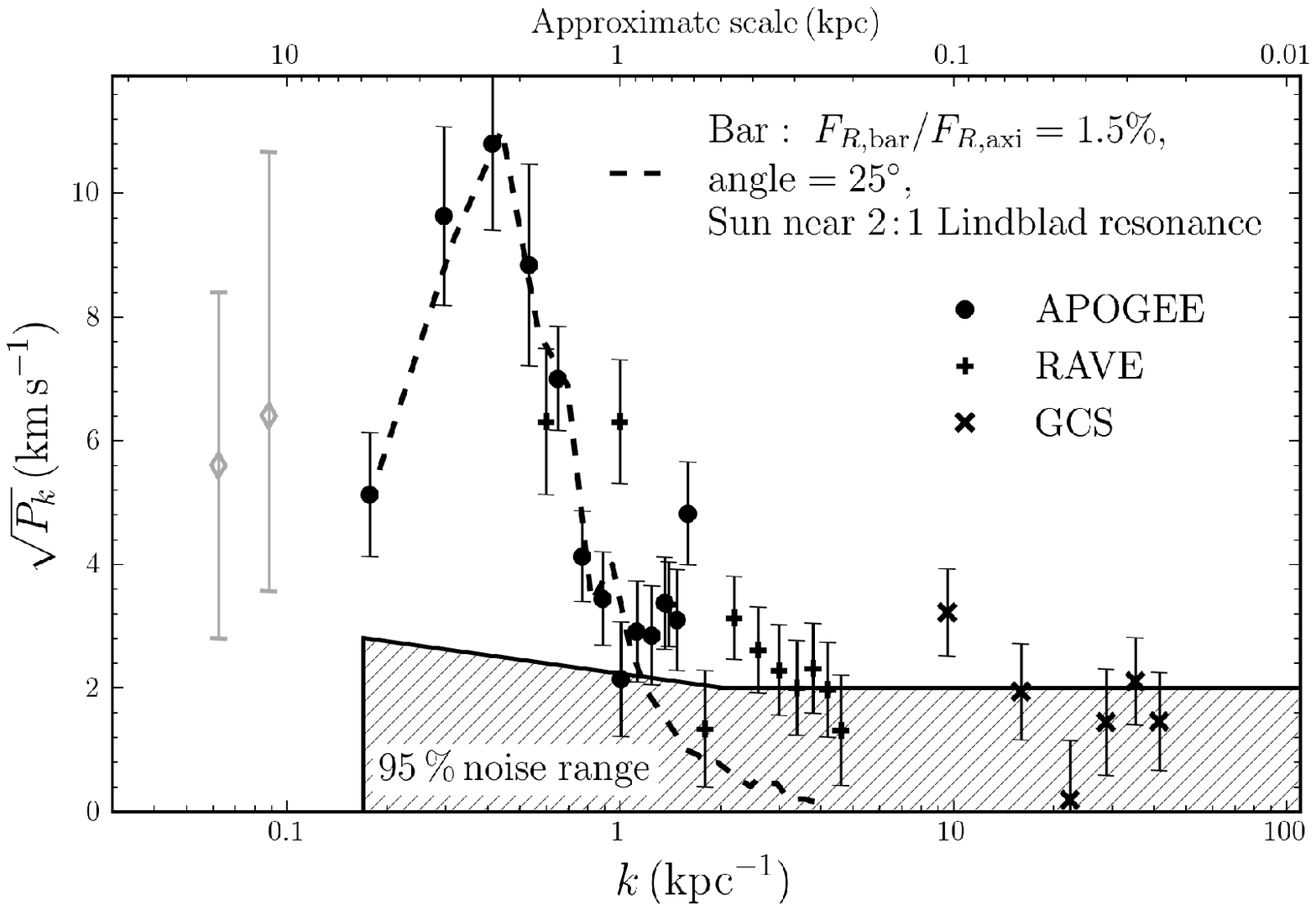}
\caption{One-dimensional power spectrum of velocity fluctuations in
  the MW disk computed from the APOGEE (filled dots), RAVE (plusses),
  and GCS (crosses) velocity fields shown in
  \figurename~\ref{fig:2dvel}. We have subtracted the median noise
  power spectrum to display the intrinsic fluctuations; the hatched
  region displays the region that contains $95\,\%$ of the noise. Most
  of the power is on scales $0.2\kpc^{-1} < k < 0.9\kpc^{-1}$.
  Inspection of the full two-dimensional power spectrum establishes
  that this power is almost entirely contained within $|\Delta k| =
  0.1$ of $(k_x,k_y) = (0.4,-0.15)$---that is, mainly radial---with a
  tail to $k_y = 0.25$ at fixed $k_x$ (all $k$ are in
  $\kpc^{-1}$). The gray points on the largest scales are estimates of
  the power on even larger scales from observations of external disk
  galaxies \citep{Rix95a}. The dashed line shows the expected power
  spectrum from the response to the central bar (see text for
  details), which can explain the peak in the power spectrum. On
  smaller scales ($k \gtrsim 1\kpc^{-1}$), the observed power spectrum
  is consistent with measurement noise.}\label{fig:velpsd}
\end{figure*}

\begin{figure}[t!]
\includegraphics[width=0.475\textwidth,clip=]{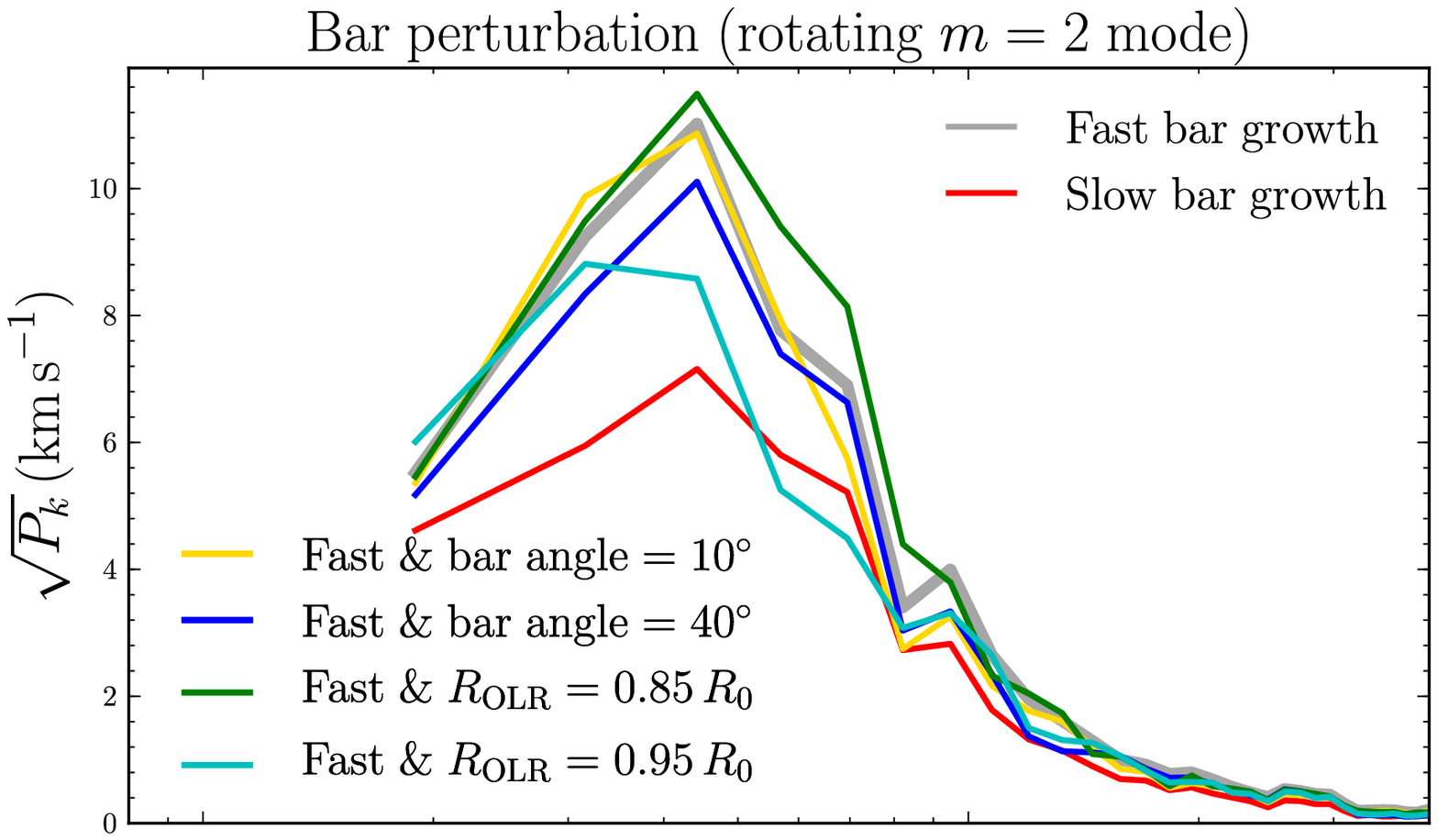}\\
\includegraphics[width=0.475\textwidth,clip=]{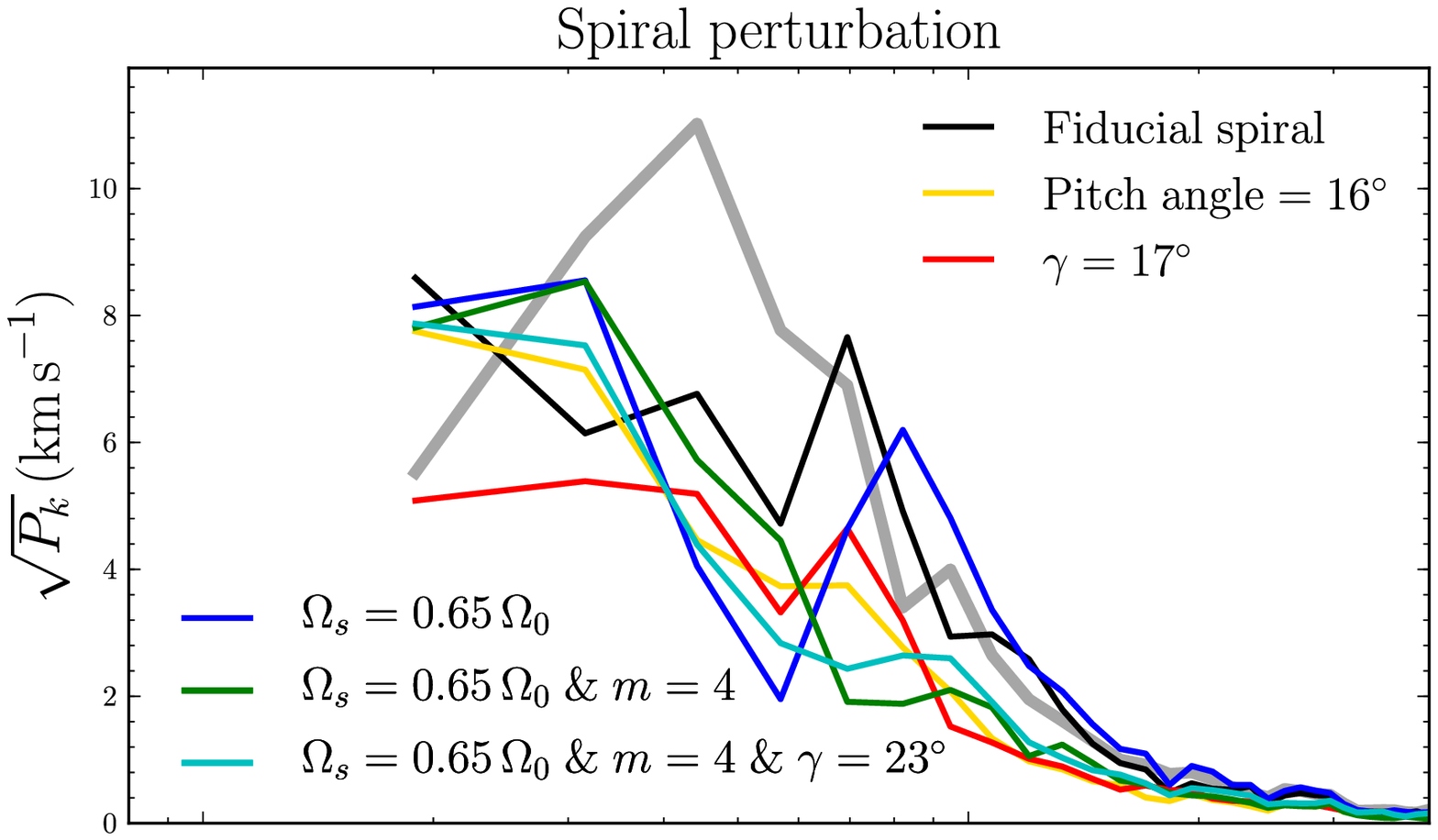}\\
\includegraphics[width=0.475\textwidth,clip=]{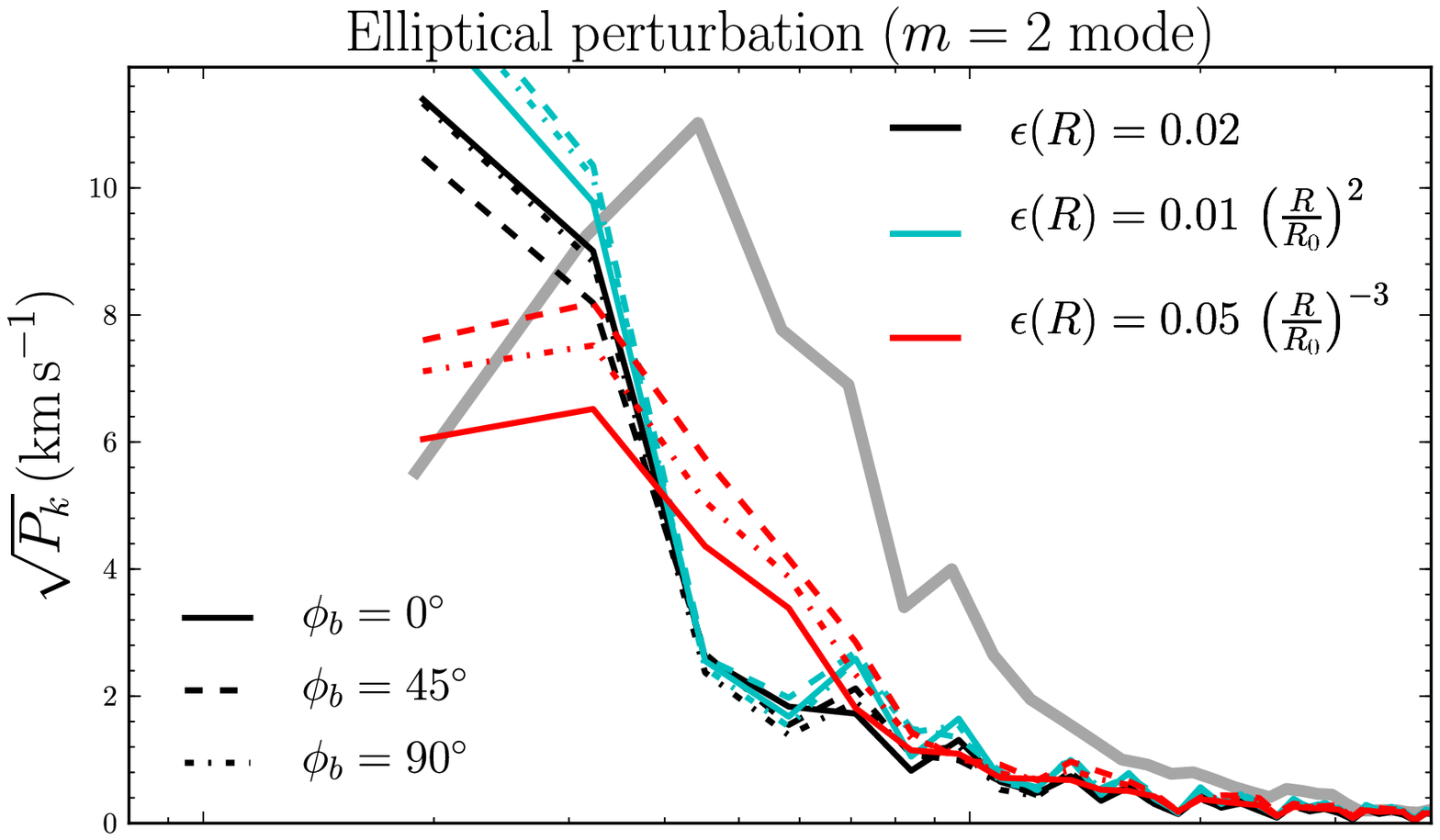}\\
\includegraphics[width=0.475\textwidth,clip=]{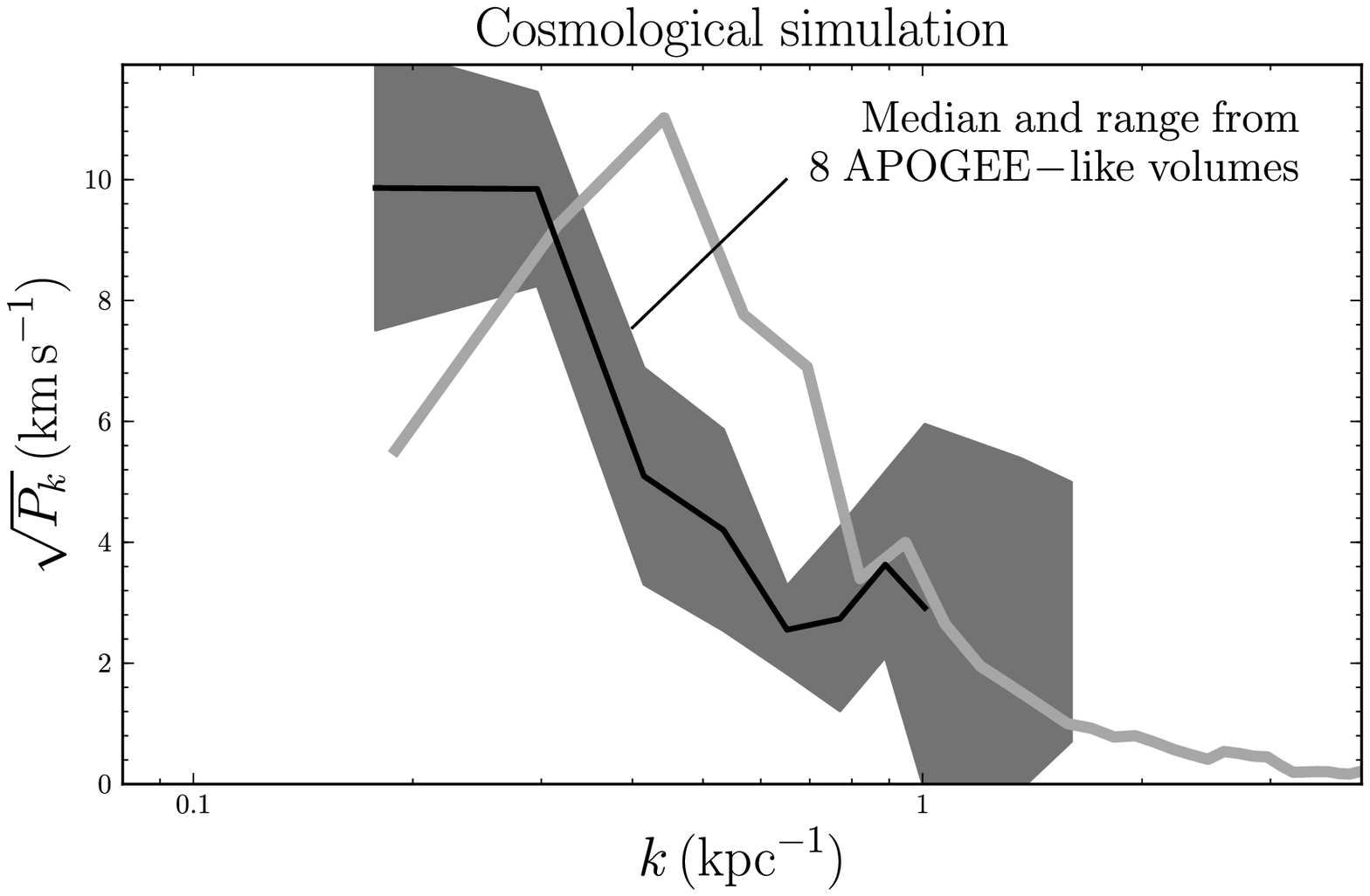}
\caption{Model power spectra for different non-axisymmetric modes: (a)
  bar modes, (b) spiral modes with various parameters, (c) elliptical
  modes, and (d) the power spectrum in the cosmological simulation
  from \citet{Guedes11a}. The various models are explained in detail
  in \sectionname~\ref{sec:modelpsd}. The fiducial bar model (thick
  gray line in all panels) fits the observed data as demonstrated in
  \figurename~\ref{fig:velpsd}. A bar is the most plausible
  explanation for the observed peak in the power spectrum in
  \figurename~\ref{fig:velpsd}.}\label{fig:velpsd_model}
\end{figure}

To determine quantitatively for which $\vsun$ the large-scale trend in
the peculiar velocity field is minimized we compute the
one-dimensional power spectrum using the method described in the
previous section. This power spectrum is discussed in more detail in
the next section. For this application, we note that most of the power
is contained in the interval $0.2\kpc\inv < k < 0.9\kpc\inv$. We can
ask for what value of $\vsun$ this power is minimized, that is, which
value of $\vsun$ minimizes large-scale power in the peculiar velocity
field. This is similar to asking what axisymmetric model for disk
kinematics best reproduces the
observations. \figurename~\ref{fig:vsolar} presents the average power
on scales $0.2\kpc\inv < k < 0.9\kpc\inv$ as a function of the assumed
$\vsun$, for different values of $V_c$ for a flat rotation curve and
for a rising and falling rotation curve with $V_c(R_0) = 220\kms$. We
also show the effect of using a different model for the stellar DF
that results in a different asymmetric drift: the thick gray line
assumes that the scale length of the exponential decline of the
velocity dispersion is $8\kpc$ (rather than the flat fiducial
profile). This model leads to an asymmetric drift that is different by
$\Delta V_a = 5\kms$ at $R_0$ and has the opposite radial dependence
(see the bottom panel of Figure 4 of B12; this model was strongly
disfavored by the data in that paper). For all of the different
assumptions about the rotation curve and stellar DF the power is
minimized for $20\kms < \vsun < 25\kms$. For a flat rotation curve
with $V_c = 220\kms$ the minimum occurs at $\vsun = 22.5\kms$.

To understand potential biases and the random uncertainty associated
with this estimate, we have performed Monte Carlo simulations of the
velocities of the RC sample using an axisymmetric Dehnen stellar DF
\citep{Dehnen99a} with $\sigma_R = 31.4\kms$, a scale length $h_R =
3\kpc$, and a flat dispersion profile in a flat-rotation curve
potential with $V_c = 220\kms$. The simulated velocity in these
simulations is generated at the catalog position of each RC star;
because we are using sub-kpc pixels and the distance uncertainties are
much smaller than this, selection and distance-error effects are
negligible for this exercise. For each of 100 simulated samples we
determine $\vsun$ in the same manner as for the data, assuming a true
$\vsun$ of $24\kms$. These simulations reveal that the estimate for
$\vsun$ is biased low by $\approx1\kms$ and has a random uncertainty
of $1\kms$. The bias is most likely due to the Gaussian approximation
for the velocity distribution, which in reality is skewed toward lower
rotational velocities. We have also performed similar Monte Carlo
simulations where the velocity field is adjusted according to the
non-axisymmetric response to the central bar (with the parameters that
we favor in \sectionname~\ref{sec:psd} below). Even in this
non-axisymmetric case, $\vsun$ as determined by the present procedure
is affected by less than $1\kms$. Therefore, our best estimate is
$\vsun = 24\pm1\kms$. A $20\kms$ higher or lower $V_c$ shifts $\vsun$
by about $2\kms$ and a similar move happens for a gently rising or
falling rotation curve, so a rather conservative systematic
uncertainty on $\vsun$ due to incomplete knowledge of $V_c(R)$ is
$2\kms$. We discuss the influence on $\vsun$ of non-axisymmetric
motions on scales larger than those observed here in
\sectionname~\ref{sec:discussion-vsolar}.

\needspace{8ex}
\section{The power spectrum of velocity fluctuations}\label{sec:psd}

\subsection{Observations}\label{sec:obspsd}

Having subtracted an axisymmetric model for the heliocentric
line-of-sight velocities in the mid-plane of the MW disk, we can now
study the peculiar velocities. On the large scales traced by APOGEE,
these are presented in the left panel of
\figurename~\ref{fig:2dvel}. The peculiar velocities have a standard
deviation of about $10\kms$, much larger than the uncertainty on the
mean peculiar velocity in each pixel, which is typically $3\kms$. To
investigate the properties of the peculiar velocity field on scales
$\lesssim 1\kpc$ that we cannot resolve well currently with APOGEE, we
use the data from RAVE and the GCS in the solar neighborhood described
in \sectionname~\ref{sec:data}. These observations are displayed in
the middle and right panel of \figurename~\ref{fig:2dvel},
respectively. For the RAVE RC stars we use the same axisymmetric model
as for APOGEE, except that we adopt $\vsun = 10\kms$, as this value
better fits the RAVE kinematics \citep{Sharma14a}. We discuss this
discrepancy between the global $\vsun$ that we employ for APOGEE and
the local value that we adopt for RAVE in much more detail in
\sectionname~\ref{sec:discussion-vsolar}. For GCS we simply plot the
median $V$ velocity from the GCS catalog, because there are no
Galactic gradients in the $\approx(100\pc)^2$ covered by the GCS
sample . This figure illustrates that the peculiar motions are mainly
confined to large scales, because the range of velocities covered by
the color scale is a factor of two and four smaller for the middle and
right panels, respectively, compared to the left panel. On the sub-kpc
scales probed by RAVE and GCS, peculiar velocities are uniformly
small.

To better quantify this behavior, we calculate the two-dimensional
power spectrum of the fluctuations shown in
\figurename~\ref{fig:2dvel}. The one-dimensional, azimuthally-averaged
power spectrum is shown in \figurename~\ref{fig:velpsd}. In detail,
this one-dimensional power spectrum is obtained by following the
procedure in \sectionname~\ref{sec:psdmethod} for the APOGEE, RAVE,
and GCS velocity fields in \figurename~\ref{fig:2dvel} separately; we
do not display the smallest $k$ and the three largest $k$ for each
data set to minimize pixelization effects. For each data set we
perform 1,000 Monte Carlo simulations of the noise by generating
Gaussian random fields scaled to the measurement uncertainty on
$\Delta V_{\mathrm{los,rot}}$ in observed pixels and we calculate
their power spectrum in the same manner. We subtract the median of
these 1,000 noise power spectra from the data and indicate the 95\,\%
noise range determined from these simulations. All power above this
95\,\% noise range is therefore detected at $2\sigma$. The combination
of the three data sets allow a measurement of the one-dimensional
power spectrum on all scales in the range $0.2\kpc\inv \leq k \leq
40\kpc\inv$, excluding only the largest spatial scales in the disk. As
an estimate of the expected power on even larger scales, we include
the typical power in elliptical and lopsided modes for MW-like
galaxies from \citet{Rix95a}. If we change the value of $\vsun$ used
for the APOGEE-RC sample or if we modify the parameters of the
asymmetric-drift model, the only change is that the height of the peak
becomes larger (reflecting the fact that the axisymmetric fit becomes
worse; see \figurename~\ref{fig:vsolar}), but the shape of the power
spectrum remains the same.  As noted by \citet{Binney14a}, the RAVE RC
distances that we employ may be underestimated by
$\approx0.08\magunit$. We have repeated the power-spectrum analysis
using $4\,\%$ larger RAVE distances and find no qualitative impact on
the observed power spectrum.

The observed power spectrum is characterized by a single large peak on
scales $0.2\kpc\inv \lesssim k \lesssim 0.9\kpc\inv$. The power on
scales smaller than $1\kpc$ ($k \gtrsim 1\kpc\inv$) is consistent with
being noise in the measurement of $\Delta V_{\mathrm{los,rot}}$. This
result is entirely expected for the kinematically-warm populations of
stars in our three samples. Any inhomogeneity on small scales would
tend to be washed out by epicyclic motions. Interestingly, the power
declines for the two smallest $k$ measured by APOGEE. This behavior
indicates that most of the power in the peculiar velocities in the MW
(at least in the region surveyed by the APOGEE-RC sample) is on the
$\approx2\kpc$ scales where the power spectrum peaks. Extrapolating
this behavior, we therefore expect that future investigations of
peculiar velocities on larger scales will find relatively little power
in, \eg, elliptical or lopsided modes. Using the lack of azimuthal
variations in the metallicity distribution in the APOGEE-RC sample as
a constraint, \citet{BovyRC} already limited elliptical modes to
streaming motions $\lesssim10\kms$, although these constraints and
equivalent constraints on lopsided modes will become significantly
better with future data from APOGEE-2, the second phase of the APOGEE
project, which will operate from 2014 to 2020 and which will
(sparsely) cover the entire disk \citep{Sobeck14a}.

An investigation of the two-dimensional power spectrum demonstrates
that most of the power is in fact in a single peak around $(k_x,k_y) =
(0.4,-0.15)\kpc\inv$, or largely in the $x$ direction. The
line-of-sight projection effects and incomplete spatial sampling make
it difficult to interpret the two-dimensional power spectrum. We only
model the one-dimensional power spectrum below, because of these
difficulties and also because the two-dimensional measurement is
noisier.

The peak in the power spectrum occurs at $k \approx 0.5\kpc\inv$,
meaning that the largest variations in the peculiar velocity field are
on $\approx2\kpc$ scales, using our frequency definitions in
\sectionname~\ref{sec:psdmethod}. Among likely culprits, such a peak
can be accurately reproduced by the response to a central bar. An
illustrative model for the response of the disk to the bar that fits
the observed peak is overlaid in \figurename~\ref{fig:velpsd}. This
bar has an angle of $25^\circ$ with respect to the
Sun--Galactic-center line, a pattern speed of $\Omega_b =
1.9\,\Omega_0$ (where $\Omega_0$ is the local rotational frequency),
and a radial-force-amplitude at $R_0$ of 1.5\% relative to the
axisymmetric force. We discuss this and alternative models further in
the next subsection.

\subsection{Modeling of non-axisymmetric perturbations}\label{sec:modelpsd}

To interpret the measurement of the power spectrum of velocity
fluctuations, we perform simulations of the effect on the velocity
field of various non-axisymmetric perturbations to the Galactic
potential, using the backward-integration technique of
\citet{Dehnen00a}. For an assumed non-axisymmetric perturbation, this
method finds the value of the stellar DF at a given phase--space point
today by reverse integration of the orbit until a time $t_0$ before
the non-axisymmetric was active; the DF today is then equal to the
axisymmetric DF at $t_0$ evaluated at the position of the orbit at
$t_0$. Specifically, we make use of the \texttt{evolveddiskdf}
implementation of this method in
\texttt{galpy}\footnote{\url{http://github.com/jobovy/galpy}~.}
\citep{BovyGalpy}. The spatial dependence of the mean velocity
response for the perturbations considered below is similar to that of
a kinematically-cold population, as expected from linear perturbation
theory \citep{binneytremaine}. However, analytically calculating the
amplitude of the kinematically-warm response is in general difficult.

All of the simulations below assume a background potential with a flat
rotation curve and an initial DF given by a Dehnen DF
\citep{Dehnen99a} with $\sigma_R = 31.4\kms$, a scale length $h_R =
3\kpc$, and a flat dispersion profile (these properties do not change
considerably due to the non-axisymmetric perturbation). We compute the
mean peculiar velocity field over $5.5\kpc \leq X_{GC} \leq 12.25\kpc$
and $-3\kpc \leq Y_{GC} \leq 3.75\kpc$ (the range of the data, see
\figurename~\ref{fig:rcdvlos}) by computing moments of the
non-axisymmetric DF and then determine its power spectrum in the same
way as for the data. The slope of the rotation curve has only a
sub-dominant effect on the power spectrum for all of the perturbations
considered below and we therefore only discuss the flat-rotation
curve.

The result of these simulations are displayed in the top two panels of
\figurename~\ref{fig:velpsd_model}. The top panel has the results for
a central bar, modeled as a simple quadrupole with the same parameters
as those of \citet{Dehnen00a}, except that the bar is 50\% stronger,
with a radial-force-amplitude at $R_0$ of 1.5\% relative to the
axisymmetric force (see the end of the previous section for the full
set of bar parameters). The gray curve demonstrates the effect of a
bar that is grown over two bar periods and subsequently evolved for
two more bar periods (the default model of \citealt{Dehnen00a}). The
red curve displays the effect of an adiabatically-grown bar; this bar
was grown over 68 bar periods and evolved for 7 more periods. The
resulting velocity power spectrum is similar for these two cases, but
the more slowly-grown bar gives rise to smaller velocity fluctuations
with a somewhat flatter spectrum on large scales. We also investigate
variations in the bar's angle and pattern speed; these do not strongly
affect the shape of the power spectrum, except for the slower bar.

The second panel in \figurename~\ref{fig:velpsd_model} demonstrates
the effect of a logarithmic spiral potential with various
parameters. The fiducial spiral model is that of an $m=2$ spiral with
a pitch angle of $8^\circ$, a pattern speed of $\Omega_s =
0.33\Omega_0$ that puts the Sun close to the $2:1$ inner Lindblad
resonance, and an angle $\gamma$ (between the Sun--Galactic-center
line and the line connecting the Galactic center to the peak of the
spiral pattern at $R_0$) of $69^\circ$. The other curves display the
effect of variations in these parameters: (a) a larger pitch angle,
(b) a different angle $\gamma$, (c) a larger pattern speed that puts
the Sun near the $4:1$ inner Lindblad resonance, (d) an $m=4$ spiral
with the latter pattern speed, and (e) case (d) with a different angle
$\gamma$. While the power spectrum of the velocity response to these
different spiral perturbations varies, it typically places most of the
power on the largest scales in a way that is inconsistent with the
observed data, such that spiral structure alone cannot explain the
observed power spectrum.

\begin{figure*}[t!]
\includegraphics[width=\textwidth,clip=]{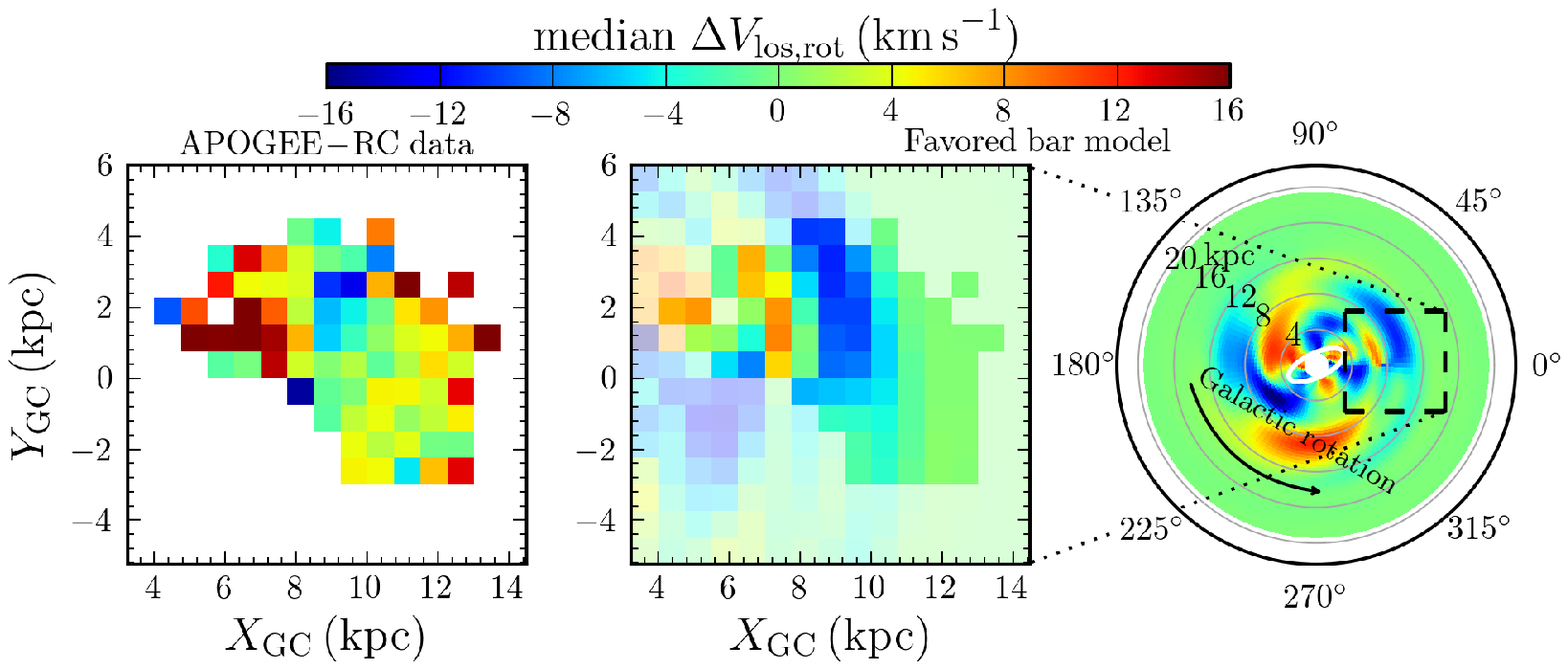}
\caption{The observed peculiar line-of-sight velocity field
  (\emph{left panel}) compared to that in the bar model favored in
  this paper. The right panel displays the peculiar line-of-sight
  velocities induced by the bar over the whole Galactic disk and the
  middle panel expands the region currently observed by APOGEE. The
  bar's orientation and extent are indicated by the white ellipse in
  the right panel. The model explains some of the large-scale features
  (positive velocities near $(X_{\mathrm{GC}},Y_{\mathrm{GC}}) =
  (7,2)\kpc$; negative velocities at $X_{\mathrm{GC}}\approx9\kpc$,
  $0< Y_{\mathrm{GC}} < 4\kpc$), but it has almost no response at $R
  \gtrsim11\kpc$. The difficulty of detecting the effect of the bar in
  the velocity field itself demonstrates why the power spectrum is a
  much more sensitive tool than the velocity field itself for
  determining the most important non-axisymmetric
  actor(s).}\label{fig:barvlos}
\end{figure*}

The third panel of \figurename~\ref{fig:velpsd_model} shows the power
spectrum of velocity fluctuations induced by an elliptical
perturbation, that is, a perturbation to the potential $\propto
\epsilon(R)\cos2[\phi-\phi_b]$ (see, \eg, \citealt{Kuijken94a}). Such
a perturbation could result from, \eg, a triaxial halo. Rather than
fully simulating the response in the manner described above, we make
use of the results of \citet{BovyGalpy}, who characterized the radial-
and rotational- velocity response of a kinematically-warm stellar disk
to an elliptical perturbation in detail using the backward-integration
technique. \citet{BovyGalpy} found that the response is well
characterized by the response of a cold disk, which can be computed
analytically, multiplied by reduction factors
$\mathcal{F}_{V_R}(\sigma_R)$ and $\mathcal{F}_{V_T}(\sigma_R)$ for
the radial and rotational velocity, respectively. The different colors
in the third panel of \figurename~\ref{fig:velpsd_model} demonstrate
the response for different radial dependencies of $\epsilon(R)$ and
the line-styles show the effect of varying the angle $\phi_b$. As
expected, regardless of the detailed form of the ellipticity, the
power spectrum always peaks on the largest scales, with little power
on the scales where the observed power spectrum peaks.

Finally, we investigate velocity fluctuations in a high-resolution
cosmological simulation. In particular, we examine ErisLE (as
described in \citealt{Bird13a}), a member of the Eris simulation suite
designed to follow the formation of Milky-Way-sized galaxies. Several
key structural and kinematic properties of this simulated galaxy,
which closely resemble those of the Milky Way at $z=0$, have been
studied extensively in \citet{Guedes11a}, \citet{Guedes13a}, and
\citet{Bird13a}. To mimic the APOGEE observations, we calculate the
mean velocity $\bar{V}(R)$ as a function of radius using all stars at
$l$ $30^\circ \leq l \leq 210^\circ$, $|b| \leq 1.5^\circ$, and height
$|z| \leq 0.25\kpc$ from a Sun-like position in the simulation (\ie,
at $8\kpc$ from the center). We compute the residuals between the
median velocity field of stars with ages between 1 and $5\Gyr$ (to
mimic the RC population) and the velocity field $\bar{V}(R)$. This
procedure is repeated for 8 different Sun-like positions. The bottom
panel of \figurename~\ref{fig:velpsd_model} displays the median power
spectrum of these 8 positions as well as the $1\sigma$ range. The
velocity fluctuations in the cosmological simulations are of a similar
magnitude as those observed in the data, but they are typically found
on the largest scales in the APOGEE-like volume. Direct inspection of
the cosmological simulation reveals that the velocity perturbations
are most likely generated by spiral structure, which is apparent in
the projected surface density, because the simulated galaxy has not
experienced a recent major merger nor is the transient, large-scale
bar prominent over the last $5\Gyr$ (the strength of the bar in ErisLE
gradually decreases during this time, likely due to a combination of
dynamical friction and gas inflow;
\citealt{Debattista06a,Guedes13a}). The power spectrum of velocity
fluctuations is consistent with this interpretation, as it is more
like that of the simple spiral simulations than that of the bar, but
it is inconsistent with the APOGEE observations.

\section{Discussion}\label{sec:discussion}

\subsection{Non-circular motions in the MW}\label{sec:discussion-noncirc}

After subtracting our best smooth, axisymmetric model for the
kinematics of APOGEE-RC stars, large peculiar velocities on scales
$\gtrsim1\kpc$ exist. Variations in the peculiar velocities on the
smaller scales observed in the RAVE and GCS samples are absent, at
least in the solar neighborhood. These results are in qualitative
agreement with the amplitude of $\approx10\kms$ of the observed
wiggles in the terminal velocities of HI/CO in the MW
\citep[\eg,][]{Levine08a} or in the peculiar velocity field in
external galaxies \citep[\eg,][]{Visser80a}, which occur on similar
scales, but have usually been interpreted as being due to spiral
structure \citep[\eg,][]{Yuan69a,Burton71a}. That there are
large-scale streaming motions at the amplitude that we observe is
therefore not surprising, but it is the first time that these motions
are measured at high signal-to-noise ratio in the intermediate--age
stellar component.

As is clear from \figurename~\ref{fig:velpsd} and
\sectionname~\ref{sec:modelpsd}, the observed power spectrum of
velocity fluctuations can be accurately modeled as being the response
to the central bar and this is the likeliest non-axisymmetric mode
that can explain the observations. A radial-force-amplitude at $R_0$
of 1.5\% relative to the axisymmetric force corresponds to a bar mass
of $\approx1.5\times10^{10}\msun$, well within the range of other
estimates of the bar mass \citep[\eg,][]{Dwek95a,Zhao96a,Weiner99a}.
\figurename~\ref{fig:barvlos} directly compares the observed
peculiar-velocity field to that in the favored bar model: the left
panel displays the observed peculiar-velocity field, the right panel
shows the model over the whole Galactic disk, and the middle panel
expands the region currently observed by APOGEE. It is clear that the
model reproduces some of the qualitative features of the observed
peculiar velocities, although the data exhibit larger fluctuations
than those due to the bar, especially at $R \gtrsim10\kpc$.

The bar model makes strong predictions for the peculiar velocity field
that can be determined from future APOGEE-2 data in the Southern
Hemisphere ($l \geq 210^\circ$) or in the outer Milky Way: at $R
\gtrsim 11\kpc$ the bar should have almost no perceptible effect on
the stellar velocity distribution. APOGEE-2 will also be able to
constrain the bar further by observations of the whole velocity
distribution rather than just its mean \citep{Bovy10a}.

That the bar is successful at explaining the observed distribution of
power in the peculiar velocity field does not necessarily mean that
the effect of other non-axisymmetric modes is negligible. We almost
certainly require additional non-axisymmetry to explain the
$\approx4\kms$ positive line-of-sight velocities at $R \gtrsim10\kpc$
in the data (see \figurename~\ref{fig:barvlos}). While it is clear
from \figurename~\ref{fig:velpsd_model} that the observed power
spectrum places tight constraints on elliptical perturbations, spiral
structure at levels that can have significant heating and migration
effects are still possible. Future investigations in regions of the
disk where the effect of the bar is small (\eg, the disk at $R \gtrsim
11\kpc$) or on much larger scales ($k \lesssim 0.2\kpc\inv$) can be
used to determine the parameters and likely effect of spiral
structure. Truly global investigations of the velocity field, however,
will be necessary for this.

\needspace{8ex}
\subsection{The global and local solar motion}\label{sec:discussion-vsolar}

Having characterized the spectrum of velocity fluctuations, we can
return to the question of why the solar motion $\vsun$ measured by
APOGEE is so much larger than that measured locally and which we
denote by $V_\odot$. Analyses of the local kinematics of stars in
\emph{Hipparcos} \citep{Dehnen98b,Hogg05a}, GCS
\citep{Binney10a,Schoenrich10a}, or RAVE \citep{Sharma14a} all find
that $V_\odot = 5$ to $12\kms$, with the latter value being considered
to be more accurate because of unmodeled correlations between the
color used to define populations of stars and the populations' radial
gradients in the former. A direct examination of the GCS catalog shows
that the local value $V_\odot$ cannot be much larger than $12\kms$,
because otherwise all of the stars in the solar neighborhood are at
their pericenters, a result that would be at odds with everything that
is known about galactic disks and the MW disk in particular. However,
the way the local $V_\odot$ is measured implies that $V_\odot$ is the
solar motion with respect to a hypothetical zero dispersion population
at the position of the Sun. If the whole solar neighborhood is
participating in a global streaming motion, then there will be an
offset between the local $V_\odot$ and the global $\vsun$, the latter
defined as the difference between the Sun's rotational velocity and
the axisymmetric $V_c$
\citep[\eg,][]{Shuter82a,Kuijken94a,Metzger98a}. This issue is
discussed in detail in B12.

Our new measurement of the peculiar velocity field proves that it is
indeed likely that the solar neighborhood is affected by a streaming
motion of $\vsun-V_\odot\approx10\kms$. The power spectrum of velocity
fluctuations demonstrates that such fluctuations are common on scales
of $\gtrsim1\kpc$, while there is little power on smaller scales. This
result directly explains why all local surveys, which are limited to
distances projected onto the plane $\lesssim1\kpc$ measure the same
$V_\odot$: They are all affected by---and are insensitive to---the
same streaming motion.

The global value of $\vsun=24\kms$ that we determined in this paper
uses the largest scales in the sample and is therefore less affected
by large-scale streaming motions. This value was first obtained by B12
using the first year of APOGEE data. Because that analysis used more
luminous and hence more distant red giants than the RC stars used
here, it determined $\vsun$ on even larger scales. The fact that the
power spectrum turns over around $k \approx 0.5\kpc\inv$ implies that
there is little power on the largest scales. Therefore, the global
value of $\vsun$ measured here and in B12 will not suffer from the
large streaming-related systematics that the local $V_\odot$
experience. Conservatively, we can assign the amplitude of the
velocity power measured at the largest scale in our
sample---$\approx5\kms$ at $k =0.16\kpc\inv$---as a systematic
uncertainty on the global value of $\vsun$.  Determinations of the
Sun's velocity on larger scales than those considered here, for
example, by modeling the kinematics of tidal streams, are also global
measurements and are by and large consistent with our measurement. For
example, the large solar velocity with respect to the Galactic center
obtained from the Sgr stream \citep{Carlin12a} is consistent with our
\vsun\ combined with $V_c \approx 220\kms$.

The streaming motion of $12\kms$ of the whole solar neighborhood must
be caused by a non-axisymmetric perturbation. The bar model favored
above to explain the RC peculiar velocity field includes a rotational
streaming motion at the Sun of $\approx4\kms$ (using
\equationname~[14] of \citealt{Muhlbauer03a}), which falls short of
explaining the observed streaming motion by $8\kms$. This discrepancy
would only be $6\kms$ if the bar angle were $10^\circ$. Nevertheless,
an additional streaming motion induced by a different non-axisymmetric
agent, such as spiral structure, is probably necessary to explain the
large streaming motion of the solar neighborhood.

With $\vsun = 24\kms$ and the measured proper motion of Sgr A$^*$ of
$30.24\kms\kpc\inv$ \citep{Reid04a},
\begin{equation}\label{eq:vc}
  V_c = 218\kms + 30.24\kms\kpc\inv\,\left(R_0 - 8\kpc\right)\,.
\end{equation}
Therefore, for $R_0 = 8\kpc$, this measurement of $V_c$ agrees
perfectly with the measurement of B12. That determination uses a
largely different stellar sample and does not include the Sgr A$^*$
proper motion, so this agreement is a genuine consistency check on
both analyses. To increase $V_c$ to $>240\kms$ requires that $R_0 >
8.7\kpc$, which is highly unlikely given the current best constraints
\citep{Ghez08a,Gillessen09a}. \Eqnname~(\ref{eq:vc}) is not entirely
correct, as \figurename~\ref{fig:vsolar} demonstrates that the $\vsun$
that we measure depends on $V_c$ in such a way that $\vsun$ is about
$2\kms$ lower if $V_c = 240\kms$. This dependence, however, only
affects the above statement by a percent.

\section{Conclusion and outlook}\label{sec:conclusion}

In this paper we have investigated the two-dimensional line-of-sight
velocity field in the MW mid-plane out to $5\kpc$ using APOGEE-RC
stars. We have focused on two important questions: (a) the Sun's
motion with respect to the circular velocity and (b) the residuals
from an axisymmetric kinematic model. We characterized the latter
using their power spectrum, finding that the power on scales in the
range $0.2\kpc\inv \leq k \leq 40\kpc\inv$ is fully contained within
$0.2\kpc\inv \leq k \leq 0.9\kpc\inv$, with the power on smaller
scales consistent with measurement noise. The most likely perturber
that creates power on these scales is the central bar.

We measured the Sun's motion with respect to the circular velocity
using a new method based on minimizing the large-scale residuals in
the peculiar velocity field. This approach unambiguously determines
$\vsun$ to be relatively large. In detail, we measure $\vsun = 24\pm
1\, (\mathrm{ran.})\pm 2\, (\mathrm{syst.}\ [V_c])\pm
5\,(\mathrm{syst.\ [large\!-\!scale]})\kms$. Here, we have included
systematic uncertainties due to (a) a $20\kms$ uncertainty in $V_c$
and (b) the estimated power on unobserved larger scales. This
measurement agrees with the determination by B12, which employed an
almost independent subsample of APOGEE data from the one used here and
which used a different technique for inferring $\vsun$.

In the future, much better measurements of the peculiar velocity field
in the MW mid-plane will be possible with data from the \emph{Gaia}
satellite \citep{Perryman01a} and from APOGEE-2. \emph{Gaia} will
allow both dimensions of the planar peculiar velocity field to be
determined, rather than just the line-of-sight component as done
here. APOGEE-2 will perform measurements at all Galactic longitudes
and at much larger distances from the Sun in the dust-obscured regions
of the disk, crucial for understanding the largest-scale modes
affecting the disk's kinematics and for a complete interpretation of
the optical \emph{Gaia} data. Performing similar measurements with
ongoing surveys of resolved kinematics in galaxies (\eg, CALIFA,
\citealt{Sanchez12a}; SAMI, \citealt{Croom12a}; MANGA,
\citealt{Bundy14a}) will be difficult because of the relatively low
spectral resolution and typical signal-to-noise ratios of these
surveys. However, MUSE \citep{Bacon10a} could be used to create maps
with high enough velocity resolution to determine the power spectrum
on the largest scales in nearby galaxies, which would provide an
interesting look into the drivers of galaxy evolution in different
types of galaxies and environments. Connecting these measurements to
that in the Milky Way will be crucial for determining the exact
characteristics and origin of the power in the velocity field on large
scales and their implications for the dynamical evolution of stellar
populations in galactic disks.



\acknowledgements J.B. gratefully acknowledges various insightful
conversations with Scott Tremaine about the MW velocity field and how
to best characterize it as well as detailed comments on this paper. It
is also a pleasure to thank Simeon Bird for discussions of two- and
one-dimensional power spectra, and Hans-Walter Rix, Annie Robin,
Donald Schneider, Matthias Steinmetz, David Weinberg, and the
anonymous referee for helpful discussions and comments. We also thank
Simone Callegari, Javiera Guedes, Piero Madau, and Lucio Mayer for
sharing the ErisLE simulation with us. J.B. was supported by NASA
through Hubble Fellowship grant HST-HF-51285.01 from the Space
Telescope Science Institute, which is operated by the Association of
Universities for Research in Astronomy, Incorporated, under NASA
contract NAS5-26555. J.B. further acknowledges support from a John
N. Bahcall Fellowship and the W.M. Keck
Foundation. J.C.B. acknowledges the support of the Vanderbilt Office
of the Provost through the Vanderbilt Initiative in Data-intensive
Astrophysics (VIDA). S.R.M. was supported by NSF grants 1109718 and
1413269.

Funding for RAVE (www.rave-survey.org) has been provided by
institutions of the RAVE participants and by their national funding
agencies.

Funding for SDSS-III has been provided by the Alfred P. Sloan
Foundation, the Participating Institutions, the National Science
Foundation, and the U.S. Department of Energy Office of Science. The
SDSS-III web site is http://www.sdss3.org/.

SDSS-III is managed by the Astrophysical Research Consortium for the
Participating Institutions of the SDSS-III Collaboration including the
University of Arizona, the Brazilian Participation Group, Brookhaven
National Laboratory, Carnegie Mellon University, University of
Florida, the French Participation Group, the German Participation
Group, Harvard University, the Instituto de Astrofisica de Canarias,
the Michigan State/Notre Dame/JINA Participation Group, Johns Hopkins
University, Lawrence Berkeley National Laboratory, Max Planck
Institute for Astrophysics, Max Planck Institute for Extraterrestrial
Physics, New Mexico State University, New York University, Ohio State
University, Pennsylvania State University, University of Portsmouth,
Princeton University, the Spanish Participation Group, University of
Tokyo, University of Utah, Vanderbilt University, University of
Virginia, University of Washington, and Yale University.


\begin{thebibliography}{}

\bibitem[Athanassoula(2003)]{Athanassoula03a}
  Athanassoula, E.\ 2003, \mnras, 341, 1179
\bibitem[Bacon \etal(2010)]{Bacon10a}
  Bacon,~R., Accardo,~M., Adjali,~L., \etal\ 2010, \procspie, 7735, 08
\bibitem[Barbanis \& Woltjer(1967)]{Barbanis67a}
  Barbanis,~B. \& Woltjer,~L. 1967, \apj, 150, 461 
\bibitem[Bensby \etal(2007)]{Bensby07a}
  Bensby,~T., Oey,~M.~S., Feltzing,~S., \& Gustafsson,~B. 2007, \apj, 655, 89
\bibitem[Binney \& Lacey(1988)]{Binney88a}
  Binney,~J. \& Lacey,~C. 1988, \mnras, 230, 597 
\bibitem[Binney \etal(1991)]{Binney91a}
  Binney,~J., Gerhard,~O.~E., Stark,~A.~A., Bally,~J., \& Uchida,~K.~I.\ 1991, \mnras, 252, 210
\bibitem[Binney, Gerhard, \& Spergel(1997)]{Binney97a}
  Binney,~J.~J., Gerhard,~O., \& Spergel,~D.~N. 1997,  \mnras, 288, 365
\bibitem[{{Binney} \& {Tremaine}(2008)}]{binneytremaine}
  Binney,~J. \& Tremaine,~S. 2008, Galactic Dynamics: Second Edition (Princeton, NJ: Princeton Univ. Press)
\bibitem[Binney(2010)]{Binney10a}
  Binney,~J. 2010, \mnras, 401, 2318
\bibitem[Binney \etal(2014)]{Binney14a}
  Binney,~J., Burnett,~B., Kordopatis,~G., \etal\ 2014, \mnras, 437, 351
\bibitem[Bird et al.(2013)]{Bird13a}
  Bird,~J.~C., Kazantzidis,~S., Weinberg,~D.~H., \etal\ 2013, \apj, 773, 43
\bibitem[Bissantz \& Gerhard(2002)]{Bissantz02a}
  Bissantz,~N.~\& Gerhard,~O. 2002, \mnras, 330, 591
\bibitem[Blitz \& Spergel(1991)]{Blitz91a}
  Blitz,~L. \& Spergel,~D.~N.\ 1991, \apj, 379, 631
\bibitem[Bournaud \etal(2009)]{Bournaud09a}
  Bournaud,~F., Elmegreen,~B.~G., \& Martig,~M. 2009, \apj, 707, 1
\bibitem[Bovy(2010)]{Bovy10a}
  Bovy,~J.\ 2010, \apj, 725, 1676
\bibitem[Bovy \& Hogg(2010)]{Bovy10b}
  Bovy,~J. \& Hogg,~D.~W.\ 2010, \apj, 717, 617
\bibitem[Bovy \etal(2012)]{BovyVc}
  Bovy,~J., Allende~Prieto,~C., Beers,~T.~C., \etal\ 2012, \apj, 759, 131 (B12)
\bibitem[Bovy \& Rix(2013)]{Bovy13a}
  Bovy,~J. \& Rix,~H.-W. 2013, \apj, 779, 115
\bibitem[Bovy \etal(2014)]{BovyRC}
  Bovy,~J., Nidever,~D.~L., Rix,~H.-W., \etal\ 2014, \apj, 790, 127
\bibitem[Bovy(2015)]{BovyGalpy}
  Bovy,~J. 2015, \apjs, in press (arXiv:1412.3451)
\bibitem[Bressan \etal(2012)]{Bressan12a}
  Bressan,~A., Marigo,~P., Girardi,~L., \etal\ 2012, \mnras, 427, 127
\bibitem[Brook et al.(2004)]{Brook04a} 
  Brook,~C.~B., Kawata,~D., Gibson,~B.~K., \& Freeman,~K.~C.\ 2004, \apj, 612, 894 
\bibitem[Bundy \etal(2015)]{Bundy14a}
  Bundy,~K., Bershady,~M.~A., Law,~D.~R., \etal\ 2015, \apj, 798, 7
\bibitem[Burton(1971)]{Burton71a}
  Burton,~W.~B. 1971, \aap, 10, 76 
\bibitem[Carlberg \& Sellwood(1985)]{Carlberg85a}
  Carlberg,~R.~G. \& Sellwood,~J.~A. 1985, \apj, 292, 79
\bibitem[Carlberg(1987)]{Carlberg87a}
  Carlberg,~R.~G. 1987, \apj, 322, 59 
\bibitem[Carlin \etal(2012)]{Carlin12a}
  Carlin,~J.~L., Majewski,~S.~R., Casetti-Dinescu,~D.~I., \etal\ 2012, \apj, 744, 25
\bibitem[Contopoulos(1980)]{Contopoulos80a}
  Contopoulos,~G. 1980, \aap, 81, 198
\bibitem[Croom et al.(2012)]{Croom12a}
  Croom,~S.~M., Lawrence,~J.~S., Bland-Hawthorn,~J., \etal\ 2012, \mnras, 421, 872
\bibitem[Debattista \& Sellwood(2000)]{Debattista00a} 
  Debattista,~V.~P. \& Sellwood,~J.~A. 2000, \apj, 543, 704
\bibitem[Debattista et al.(2006)]{Debattista06a} 
  Debattista,~V.~P., Mayer,~L., Carollo,~C.~M., \etal\ 2006, \apj, 645, 209
\bibitem[Dehnen(1998)]{Dehnen98a}
  Dehnen,~W. 1998, \aj, 115, 2384
\bibitem[Dehnen \& Binney(1998)]{Dehnen98b}
  Dehnen,~W. \& Binney,~J.~J.\ 1998, \mnras, 298, 387
\bibitem[Dehnen(1999)]{Dehnen99a}
  Dehnen,~W.\ 1999, \aj, 118, 1201
\bibitem[Dehnen(2000)]{Dehnen00a}
  Dehnen,~W. 2000, \aj, 119, 800
\bibitem[De Simone et al.(2004)]{DeSimone04a}
  De Simone,~R., Wu,~X., \& Tremaine,~S. 2004, \mnras, 350, 627
\bibitem[Dwek et al.(1995)]{Dwek95a}
  Dwek,~E., Arendt,~R.~G., Hauser,~M.~G., \etal\ 1995, \apj, 445, 716 
\bibitem[Eisenstein \etal(2011)]{Eisenstein11a}
  Eisenstein,~D.~J., Weinberg,~D.~H., Agol,~E., \etal\ 2011, \aj, 142, 72
\bibitem[ESA (1997)]{ESA97a}
  ESA 1997, The \emph{Hipparcos} and Tycho Catalogues (Noordwijk: ESA: ESA
  SP-1200)
\bibitem[Famaey \etal(2008)]{Famaey08a}
  Famaey,~B., Siebert,~A., \& Jorissen,~A. 2008, \aap, 483, 453
\bibitem[Faure \etal(2014)]{Faure14a}
  Faure,~C., Siebert,~A., \& Famaey,~B. 2014, \mnras, 440, 2564
\bibitem[Ghez \etal(2008)]{Ghez08a}
  Ghez,~A.~M., Salim,~S., Weinberg,~N.~N., \etal\ 2008, \apj, 689, 1044
\bibitem[Gillessen \etal(2009)]{Gillessen09a}
  Gillessen,~S., Eisenhauer,~F., Trippe,~S., Alexander,~T., Genzel,~R., Martins,~F., Ott,~T.\ 2009, \apj, 692, 1075
\bibitem[Guedes et al.(2011)]{Guedes11a}
  Guedes,~J., Callegari,~S., Madau,~P., \& Mayer,~L.\ 2011, \apj, 742, 76 
\bibitem[Guedes \etal(2013)]{Guedes13a} 
  Guedes,~J., Mayer,~L., Carollo,~M., \& Madau,~P. 2013, \apj, 772, 36
\bibitem[Gunn \etal(2006)]{Gunn06a}
  Gunn,~J.~E., Siegmund,~W.~A., Mannery,~E.~J. \etal\ 2006, \aj, 131, 2332
\bibitem[H{\"a}nninen \& Flynn(2002)]{Hanninen02a}
  H{\"a}nninen,~J. \& Flynn,~C. 2002, \mnras, 337, 731 
\bibitem[Hogg \etal(2005)]{Hogg05a}
   Hogg,~D.~W., Blanton,~M.~R., Roweis,~S.~T., \& Johnston,~K.~V. 2005, \apj, 629, 268
\bibitem[Ida et al.(1993)]{Ida93a}
  Ida,~S., Kokubo,~E. \& Makino,~J. 1993, \mnras, 263, 875 
\bibitem[Jenkins \& Binney(1990)]{Jenkins90a}
  Jenkins,~A. \& Binney,~J. 1990, \mnras, 245, 305 
\bibitem[Kordopatis \etal(2013)]{Kordopatis13a}
  Kordopatis,~G., Gilmore,~G., Steinmetz,~M., \etal\ 2013, \aj, 146, 134
\bibitem[Kuijken \& Tremaine(1994)]{Kuijken94a}
  Kuijken~K. \& Tremaine,~S.\ 1994, \apj, 421, 178
\bibitem[Lacey \& Ostriker(1985)]{Lacey85a}
  Lacey,~C.~G. \& Ostriker,~J.~P. 1985, \apj, 299, 633
\bibitem[Laney \etal(2012)]{Laney12a}
  Laney,~C.~D., Joner,~M.~D., \& Pietrzy{\'n}ski,~G. 2012, \mnras, 419, L1637
\bibitem[Levine \etal(2008)]{Levine08a}
  Levine,~E.~S., Heiles,~C., \& Blitz,~L.\ 2008, \apj, 679, 1288
\bibitem[Majewski \etal(2011)]{Majewski11a}
  Majewski,~S.~R., Zasowski,~G., \& Nidever,~D.~L.\ 2011, \apj, 739, 25
\bibitem[Metzger et al.(1998)]{Metzger98a} 
  Metzger,~M.~R., Caldwell,~J.~A.~R., \& Schechter, P.~L. 1998, \aj, 115, 635
\bibitem[Minchev \& Famaey(2010)]{Minchev10a}
  Minchev,~I. \& Famaey,~B.\ 2010, \apj, 722, 112 
\bibitem[M{\"u}hlbauer \& Dehnen(2003)]{Muhlbauer03a}
  M{\"u}hlbauer,~G. \& Dehnen, W. 2003, \aap, 401, 975
\bibitem[Nordstr{\"o}m \etal(2004)]{Nordstroem04a}
  Nordstr{\"o}m,~B., Mayor,~M., Andersen,~J., \etal\ 2004, \aap, 418, 989 
\bibitem[Perryman \etal(2001)]{Perryman01a}
  Perryman,~M.~A.~C., de Boer,~K.~S., Gilmore,~G., \etal\ 2001, \aap, 369, 339
\bibitem[Press \etal(2007)]{Press07a}
  Press,~W.~H., Teukolsky,~S.~A, Vetterling,~W.~T., \& Flannery,~B.~P., 2007,
  Numerical Recipes: The Art of Scientific Computing, 3rd Edition (Cambridge University Press)
\bibitem[Quinn \etal(1993)]{Quinn93a}
  Quinn,~P.~J., Hernquist,~L., \& Fullagar,~D.~P.\ 1993, \apj, 403, 74
\bibitem[Reid \& Brunthaler(2004)]{Reid04a}
  Reid,~M.~J. \& Brunthaler,~A.\ 2004, \apj, 616, 872
\bibitem[Rix \& Zaritsky(1995)]{Rix95a}
  Rix,~H.-W. \& Zaritsky,~D. 1995, \apj, 447, 82
\bibitem[S{\'a}nchez \etal(2012)]{Sanchez12a}
  S{\'a}nchez,~S.~F., Kennicutt,~R.~C., Gil de Paz,~A., \etal\ 2012, \aap, 538, A8 
\bibitem[Sch\"{o}nrich \etal(2010)]{Schoenrich10a}
  Sch\"{o}nrich,~R., Binney,~J.~J., \& Dehnen,~W.\ 2010, \mnras, 403, 1829
\bibitem[Sellwood \& Binney(2002)]{Sellwood02a}
  Sellwood,~J.~A.~\& Binney,~J.J. 2002, \mnras, 336, 785
\bibitem[Sellwood(2010)]{Sellwood10a}
  Sellwood,~J.~A. 2010, \mnras, 409, 145
\bibitem[Siebert et al.(2011)]{Siebert11a}
  Siebert,~A., Famaey,~B., Minchev,~I., \etal\ 2011, \mnras, 412, 2026
\bibitem[Sharma \etal(2014)]{Sharma14a} 
  Sharma,~S., Bland-Hawthorn,~J., Binney,~J., \etal\ 2014, \apj, 793, 51
\bibitem[Shuter(1982)]{Shuter82a}
  Shuter,~W.~L.~H.\ 1982, \mnras, 199, 109
\bibitem[Sobeck \etal(2014)]{Sobeck14a}
  Sobeck,~J., Majewski,~S., Hearty,~F., \etal\ 2014, in American Astronomical Society Meeting Abstracts, 223, \#440.06
\bibitem[Spitzer \& Schwarzschild(1951)]{Spitzer51a}
  Spitzer,~L.,~Jr. \& Schwarzschild,~M. 1951, \apj, 114, 385
\bibitem[Steinmetz \etal(2006)]{Steinmetz06a}
  Steinmetz,~M., Zwitter,~T., Siebert,~A., \etal\ 2006, \aj, 132, 1645
\bibitem[Stinson et al.(2013)]{Stinson13} 
  Stinson,~G.~S., Bovy,~J., Rix,~H.-W., \etal\ 2013, \mnras, 436, 625 
\bibitem[Toth \& Ostriker(1992)]{Toth92a}
  Toth,~G. \& Ostriker,~J.~P. 1992, \apj, 389, 5
\bibitem[Velazquez \& White(1999)]{Velazquez99a}
  Velazquez,~H. \& White,~S.~D.~M. 1999, \mnras, 304, 254
\bibitem[Visser(1980)]{Visser80a}
  Visser,~H.~C.~D. 1980, \aap, 88, 149
\bibitem[Weinberg(1985)]{Weinberg85a}
  Weinberg,~M.~D. 1985, \mnras, 213, 451
\bibitem[Weiner \& Sellwood(1999)]{Weiner99a}
  Weiner,~B.~J. \& Sellwood,~J.~A.\ 1999, \apj, 524, 112 
\bibitem[Wielen(1977)]{Wielen77a}
  Wielen, R. 1977, \aap, 60, 263
\bibitem[Williams et al.(2013)]{Williams13a}
  Williams,~M.~E.~K., Steinmetz,~M., Binney,~J., \etal\ 2013, \mnras, 436, 101
\bibitem[Wilson \etal(2010)]{Wilson10a}
  Wilson,~J.~C., Hearty,~F., Skrutskie,~M.~F., \etal\ 2010, Proc.~SPIE, 7735, 46
\bibitem[Yuan(1969)]{Yuan69a}
  Yuan,~C. 1969, \apj, 158, 871 
\bibitem[Zasowski \etal(2013)]{Zasowski13a}
  Zasowski,~G., Johnson,~J.~A., Frinchaboy,~P.~M., \etal\ 2013, \aj, 146, 81
\bibitem[Zhao \& Mao(1996)]{Zhao96a}
  Zhao,~H. \& Mao,~S.\ 1996, \mnras, 283, 1197
\end{thebibliography}
\end{document}